\documentclass[aps,
prl,
reprint,
floatfix,
]{revtex4-2}

\usepackage{titletoc}
\usepackage[titles]{tocloft}
\usepackage{changepage}

\usepackage{graphicx}
\usepackage{physics}
\usepackage{booktabs}
\usepackage{amsthm,amssymb,mathrsfs}
\usepackage{tensor}
\usepackage{bm}
\usepackage[caption=false]{subfig}

\usepackage{hyperref} 
\hypersetup{colorlinks=true,allcolors=blue}

\usepackage{xargs} 
\usepackage{xcolor}

\usepackage[inline]{enumitem}

\let\temp\phi
\let\phi\varphi
\let\varphi\temp

\newcommand{\iu}{\mathrm{i}} % imaginary unit
\newcommand{\e}{\mathrm{e}} % exponent

\newcommand{\DD}{\mathscr{D}} % Path integral measure
\newcommand{\D}{\mathrm{d}} % Differential
\newcommand{\sign}{\operatorname{sgn}} % sign function
\providecommand{\ZZ}{\mathbb{Z}} % Integers
\newcommand{\normord}[1]{\ensuremath{{\bm{:}}\mathbin{#1}{\bm{:}}}}

\let\v\temp % 
\let\temp\vv
\let\v\relax
\newcommand{\v}[1]{\ensuremath{\mathbf{#1}}} % vector
\makeatletter
\newcommand*{\coloneqq}{\mathrel{\rlap{%
			\raisebox{0.28ex}{$\m@th\cdot$}}%
		\raisebox{-0.28ex}{$\m@th\cdot$}}%
	=}
\newcommand*{\eqqcolon}{=\mathrel{\rlap{%
			\raisebox{0.28ex}{$\m@th\cdot$}}%
		\raisebox{-0.28ex}{$\m@th\cdot$}}%
}
\makeatother

\makeatletter
\newcommand*{\rom}[1]{\expandafter\@slowromancap\romannumeral #1@}
\makeatother

\begin{document}
	\title{Divergent Spin Conductivity on the Verge of Ferromagnetic Quantum Criticality}
	
	\author{Sondre Duna Lundemo} 
	\affiliation{Center for Quantum Spintronics, Department of Physics, Norwegian University of Science and Technology, NO-7491 Trondheim, Norway}
	
	\author{Asle Sudb\o}
	\email[Corresponding author: ]{asle.sudbo@ntnu.no}
	\affiliation{Center for Quantum Spintronics, Department of Physics, Norwegian University of Science and Technology, NO-7491 Trondheim, Norway}
	
	\date{\today} 
	
	\begin{abstract}
        We show that the spin conductivity of a metal approaching a ferromagnetic quantum critical point exhibits divergent fluctuation corrections. 
        This effect arises from critical spin fluctuations and constitutes a spin analog of the Aslamazov-Larkin theory of paraconductivity in superconductors.  
        The spin current is derived in linear response within a Gaussian-level treatment of the effective action for a system with easy-plane magnetic anisotropy. 
        We demonstrate the consistency of our spin transport theory by showing that it (i) fulfills the Ward identity and (ii) yields vanishing spin stiffness in the normal state.
        The critical enhancement of the spin conductivity is interpreted as incipient spin superfluidity in the quantum critical region.
        This is further supported by a physical picture based on the current-loop representation of the easy-plane ferromagnet.
	\end{abstract}
	
	\maketitle 
	
	\paragraph{Introduction.}\label{sec:intro}

    Quantum criticality lies at the heart of contemporary condensed matter physics, shaping our current understanding of quantum phases of matter \cite{Sachdev-Sachdev-QuantumPhaseTransitions-2011}.
    The earliest model within this paradigm was conceived by Stoner \cite{Stoner-Stoner-CollectiveElectronSpecific-1936,Stoner-Stoner-CollectiveElectronFerromagnetism-1938} who introduced a theory for a quantum critical point (QCP) in an itinerant magnet, subsequently studied in more modern language by Hertz \cite{Hertz-Hertz-QuantumCriticalPhenomena-1976} and Millis \cite{Millis-Millis-EffectNonzeroTemperature-1993}.
    At this QCP, the order-parameter fluctuations develop correlations that extend to infinite spatial and temporal scales.
    These can have dramatic consequences for electronic properties and even the stability of the Fermi liquid itself \cite{Belitz-Vojta-TransportAnomaliesMarginalFermiLiquid-2000,Chubukov-Chubukov-SelfgeneratedLocalityFerromagnetic-2005}.
    Despite its apparent simplicity, the model is anything but trivial, displaying unconventional properties at the mean-field level and beyond \cite{Belitz-Vojta-NonanalyticBehaviorSpin-1997,Belitz-Kirkpatrick-QuantumCriticalBehaviour-1996,Belitz-Vojta-FirstOrderTransitions-1999,Chubukov-Rech-InstabilityQuantumCriticalPoint-2004,Rech-Chubukov-QuantumCriticalBehavior-2006,Brando-Kirkpatrick-MetallicQuantumFerromagnets-2016,Raines-Chubukov-TwodimensionalStonerTransitions-2024,Raines-Chubukov-UnconventionalDiscontinuousTransitions-2024,Mayrhofer-Chubukov-StonerTransitionFinite-2025}.

    Many interesting phenomena occur close to these QCPs, where quantum fluctuations precede the ordering by leaving observable signatures in transport properties that reflect the universality of the QCP and the nature of the ordered phase.  
    One of the first classical examples of such an effect was provided by the seminal works of Aslamazov and Larkin \cite{Aslamazov-Larkin-InfluenceFluctuationPairing-1968,Aslamazov-Larkin-EffectFluctuationsProperties-1968}, Maki \cite{Maki-Maki-CriticalFluctuationOrder-1968}, and Thompson \cite{Thompson-Thompson-MicrowaveFluxFlow-1970}, each offering new insights into how superconducting fluctuations above $T_c$ affect the conductivity. 
    Similar fluctuation effects on the conductivity have been considered in itinerant magnets \cite{Paul-Maslov-QuantumCorrectionConductivity-2005,Paul-Paul-InteractionCorrectionConductivity-2008,Zala-Aleiner-InteractionCorrectionsIntermediate-2001}. 
    However, since the ordered phase is not characterized by dissipationless charge transport in this case, the conductivity does not receive a critical enhancement as the transition is approached.  
    In a system with easy-plane magnetic anisotropy, we may adopt an interpretation of the ordered state as a \textit{spin superfluid}, referring to the coherent spin transport mediated by topologically stable configurations of the order parameter in easy-plane magnets \cite{Halperin-Hohenberg-HydrodynamicTheorySpin-1969,Sonin-Sonin-AnalogsSuperfluidCurrents-1978,Sonin-Sonin-SuperflowsSuperfluidity-1982,Nogueira-Bennemann-SpinJosephsonEffect-2004,Sonin-Sonin-SpinCurrentsSpin-2010,Sonin-Sonin-SpinSuperfluidityCoherent-2013,Sonin-Sonin-SpinMassSuperfluidity-2018,Zhu-Tserkovnyak-ProposalSpinSuperfluid-2025}.
    Based on this perspective, it is expected that quantum spin fluctuations strongly enhance the spin conductivity in the immediate vicinity of the QCP, analogous to the enhancement of charge conductivity by fluctuating pairs close to the superconducting critical point \cite{Aslamazov-Larkin-EffectFluctuationsProperties-1968,Aslamazov-Larkin-InfluenceFluctuationPairing-1968,Maki-Maki-CriticalFluctuationOrder-1968,Thompson-Thompson-MicrowaveFluxFlow-1970}.
    
    In this letter, we unravel this phenomenon in the easy-plane Stoner model.
    Following the Gaussian-fluctuation approximation, we identify the consistent set of fluctuation diagrams that produce a physical theory of spin transport in itinerant magnets.
    The most singular corrections to the spin conductivity are computed, establishing a possible new transport signature of quantum criticality. 
    Specifically, our theory provides a potentially useful diagnostic tool for the precursor to quantum critical behavior in itinerant magnets.

    \begin{figure}[htb]
        \centering
        \includegraphics[width=\linewidth]{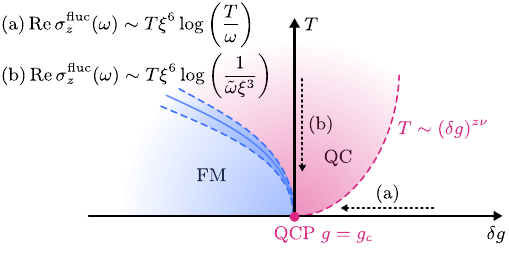}
        \caption{Schematic illustration of two ways to approach the QCP (dotted lines): (a) at low temperatures tuning $\delta g \coloneqq g_c - g \to 0^{+}$ where $\xi^{-2} \sim \delta g$, or (b) at $g=g_c$ lowering the temperature $T\to 0^{+}$ where $\xi^{-1} \sim T/\epsilon_F$. QC denotes the quantum critical region and FM the ferromagnetic metal. The dashed lines separate the quantum and classical critical regimes \cite{Altland-Simons-CondensedMatterField-2023}. The low-frequency fluctuation correction to the spin conductivity is denoted by $\Re\sigma_{z}^{\mathrm{fluc}}(\omega)$ and $\tilde{\omega} \sim \omega/\epsilon_F$.}
        \label{fig:schematic_pd}
    \end{figure}

    \paragraph{Gaussian-fluctuation action.}\label{sec:action}

    Here we consider a Fermi liquid in three dimensions with quadratically dispersing quasiparticles $\xi(\v{k}) = \v{k}^2 /(2m) - \mu$ that interact through a ferromagnetic XY exchange interaction
    \begin{equation}
        \hat{H}_{\mathrm{int}} = - \frac{1}{2} \int \D^3 r \int \D^3 r' J(\v{r}-\v{r}') \normord{S^{+}(\v{r}) S^{-}(\v{r}')},
    \end{equation}
    where $S^{\pm}(\v{r}) \coloneqq S^{x}(\v{r}) \pm \iu S^{y}(\v{r})$ and $S^{\alpha}(\v{r}) \coloneqq c^{\dagger}(\v{r}) \sigma^{\alpha} c(\v{r})/2$ is the spin-density operator.
    For simplicity, we will assume a contact interaction $J(\v{r}-\v{r}') = J\delta(\v{r}-\v{r}')$.
    The dimensionless coupling relevant for the Stoner instability is the product of the interaction strength $J/2$ and the density of states at the Fermi level, and is denoted by $g \coloneqq J \nu(0)/2$.
    In the following we will use the four-vector notation $x^{\mu} \coloneqq (-\iu\tau,\v{r})$ and $k^{\mu} \coloneqq (\iu \omega_n, \v{k})$, as well as the short-hand notation $\int \D x \coloneqq \int \D \tau \int \D^3 r$. We use units in which $\hbar = k_B = 1$.
    
    We assume that the system is subjected to an easy-plane anisotropy, reducing the global spin-rotational invariance to $\mathrm{U}(1)_z$.
    Neglecting spin-orbit coupling and disorder altogether allows for an unambiguous definition of the Noether spin current \cite{Rashba-Rashba-SpinCurrentsThermodynamic-2003,Sonin-Sonin-EquilibriumSpinCurrents-2007,Tokatly-Tokatly-EquilibriumSpinCurrents-2008,Droghetti-Tokatly-SpinorbitInducedEquilibrium-2022}. 
    The global symmetry is gauged by introducing a ``spin gauge field" $A_{\mu}$ that will serve a purpose analogous to that of the gauge field in electromagnetic linear response theory.
    The partition function in the presence of this gauge field is given by the Hubbard-Stratonovich (HS) functional integral \cite{--SeeSupplementalMaterial-} 
    \begin{equation}\label{eq:Z[A]_general}
        Z[A] = \int \DD[\bm{\pi}] \e^{ - \int \D x \frac{\bm{\pi}^2(x)}{2J} + \tr \log \left( - \beta \mathcal{G}^{-1}[A,\bm{\pi}]\right)},
    \end{equation}
    where $\mathcal{G}^{-1}[A,\bm{\pi}] = \mathcal{G}^{-1}_{0}[A] - \Sigma[\bm{\pi}]$ consists of the inverse of the bare Green's function $\mathcal{G}_{0}[A] = G_{0}[A] \bm{1}$ and the self energy $\Sigma[\bm{\pi}](x,y) = - \delta(x-y) \bm{\pi}(x) \cdot \bm{\sigma}/2$ where $\bm{\pi} \coloneqq (\pi_{x},\pi_{y})$ is the auxiliary HS field representing the in-plane magnetization.

    In the Gaussian-fluctuation approximation, the auxiliary HS field is expressed as $\bm{\pi}(x)= \bm{\pi}_{\mathrm{MF}} + \bm{\phi}(x)$ and the action appearing in Eq.~\eqref{eq:Z[A]_general} is expanded to quadratic order in fluctuations $\bm{\phi}$ about a saddle-point configuration $\bm{\pi}_{\mathrm{MF}}$. 
    This approximation ensures tractability, since we can do the remaining functional integral in Eq.~\eqref{eq:Z[A]_general}.
    In the Hertz-Millis framework it is moreover justified by the fact that the upper critical dimension is $d_c = 1 < 3$.
    Assuming an expansion about the normal state, we find 
    \begin{equation}\label{eq:Seff[A]}
        S_{\mathrm{eff}}[A] = - \tr \log\left( - \beta \mathcal{G}_{0}^{-1}[A]\right) + \tr \log \left(-\beta D^{-1}[A]\right),
    \end{equation}
    where $D^{-1}[A] = - (1 + J \Pi[A] /2  )/(2J)$ denotes the spin-fluctuation propagator and $\Pi[A]$ is the particle-hole bubble in the presence of the gauge field $A_{\mu}$. 
    In the absence of $A_{\mu}$ and near the Stoner instability defined by $ g = 1 $, the spin-fluctuation propagator can be expressed as \cite{Hertz-Hertz-QuantumCriticalPhenomena-1976} 
    \begin{equation}\label{eq:D_propagator_inverse}
        D^{-1}(p) = -\frac{1}{2J} \left( \xi^{-2} + \frac{g}{3} \hat{\v{p}}^2 + \frac{g \pi}{2} \frac{|\hat{p}_0|}{|\hat{\v{p}}|}  \right),
    \end{equation}
    where $\hat{\v{p}} \coloneqq \v{p} /(2 k_F)$, $\hat{p}_0 \coloneqq \iu \Omega_{m}/4 \epsilon_F$ and $\xi^{-2} = 1 - g$ is the zero-temperature deviation from the QCP.   
    At finite temperatures, one can approach the QCP by tuning $g = g_c$ and lowering the temperature, so that $ \xi^{-1} \sim T/\epsilon_F$. 
    The two ways of approaching the QCP are illustrated in Fig.~\ref{fig:schematic_pd}.
    
    \paragraph{Fluctuation response kernel.}\label{sec:response}

    \begin{figure*}[htb]
        \centering
        \includegraphics[width=\linewidth]{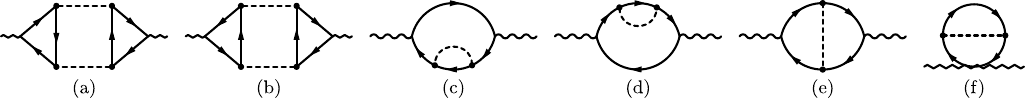}
        \caption{Fluctuation corrections to the response kernel. Figs.~(a) and (b) are the two inequivalent AL diagrams, (c) and (d) are the DOS diagrams, (e) is the MT diagram and (f) is the DIA diagram.
        The dashed lines are the spin-fluctuation propagators, the plain lines are fermion propagators and the external wiggly lines are the external spin gauge field.}
        \label{fig:diagrams}
    \end{figure*}

    With an effective action $S_{\mathrm{eff}}[A] = - \log Z[A]$ at hand, one can probe how the system responds to a weak perturbation in $A_{\mu}$ by computing the response kernel \cite{Altland-Simons-CondensedMatterField-2023}.  
    This linear response theory relates the spin-$z$ current $J_{z}^{\mu}$ to the external gauge field $A_{\mu}$ as
    \begin{equation}\label{eq:J_general}
        J^{\mu}_{z}(x) = \int \D x' Q^{\mu\nu}(x,x') A_{\nu}(x').
    \end{equation}
    We restrict our focus to the fluctuation response kernel $Q^{\mu\nu}_{\mathrm{fluc}}(x,x')$ which is obtained by performing two functional derivatives of the Gaussian-fluctuation part of the action in Eq.~\eqref{eq:Seff[A]}
    \begin{equation}\label{eq:Q_fluc}
        Q^{\mu\nu}_{\mathrm{fluc}}(x,x') = \frac{\delta^2 \tr \log \left(- \beta D^{-1}[A]\right)}{\delta A_{\mu}(x) \delta A_{\nu}(x')} \biggr\lvert_{A = 0}.
    \end{equation}
    This yields 
    \begin{subequations}
        \begin{align}
            \begin{split}
                Q^{\mu\nu}_{\mathrm{fluc}}(x,x') &= - \int \prod_{i=1}^{4} \D z_{i} \,  D(z_1,z_2) \Lambda^{\mu}(z_2,z_3;x) \label{eq:AL_realspace} \\
            &\hspace{5.1em} \times D (z_3,z_4) \Lambda^{\nu}(z_4,z_1;x') 
            \end{split} \\
        &+ \int \prod_{i=1}^{2} \D z_{i} \, D(z_1,z_2) \Xi^{\mu\nu}(z_2,z_1;x,x'), \label{eq:MT_DOS_DIA_realspace}
    \end{align}
    \end{subequations}
    where the effective vertices are given by
    \begin{subequations}
        \begin{align}
            \Lambda^{\mu}(z_1,z_2;x) &\coloneqq \frac{\delta D^{-1}[A](z_1,z_2)}{\delta A_{\mu}(x)} \biggr\lvert_{A = 0} \label{eq:three_pt_realspace}\\
            \Xi^{\mu\nu}(z_1,z_2;x,x') &\coloneqq \frac{\delta^2 D^{-1}[A](z_1,z_2)}{\delta A_{\mu}(x) \delta A_{\nu}(x')} \biggr\lvert_{A = 0}. \label{eq:four_pt_realspace}
        \end{align}
    \end{subequations}
    The fluctuation corrections to the response kernel are represented by the six diagrams in Fig.~\ref{fig:diagrams}.
    Adopting the common terminology from the corresponding diagrams in the literature on superconducting fluctuations, one obtains the Aslamazov-Larkin (AL) diagrams from Eq.~\eqref{eq:AL_realspace}, while the self-energy, or ``density-of-states" (DOS) diagrams, Maki-Thompson (MT) diagram and the diamagnetic (DIA) diagram derive from Eq.~\eqref{eq:MT_DOS_DIA_realspace}.

    The diagrams shown in Fig.~\ref{fig:diagrams} constitute the complete set of fluctuation contributions to the spin-current response kernel in the Gaussian-fluctuation approximation.
    That these diagrams should be treated on equal footing is not obvious from their diagrammatic representations \cite{Lundemo-Sudbo-FluctuationConductivityUltraclean-2026b}.
    However, the functional-integral derivation demonstrates transparently that these diagrams arise from a consistent approximation of the effective action $S_{\mathrm{eff}}[A] = - \log Z[A]$.
    This level of rigor was first introduced in the theory of superconducting fluctuations by Ref.~\cite{Boyack-Boyack-RestoringGaugeInvariance-2018}, who revealed a missing diagram (the analog of Fig. \ref{fig:diagrams} (f)) of the original theory which was crucial to obtain a physically meaningful transport theory.  
    In the Supplemental material \cite{--SeeSupplementalMaterial-} we show that the present theory also meets the same standards; it is proved that the sum of these diagrams ensures the gauge invariance of the effective action $S_{\mathrm{eff}}[A]$ and that the superfluid stiffness vanishes in the normal state. 
    These constraints translate to $q_{\mu} Q^{\mu\nu}_{\mathrm{fluc}}(q) = 0$ and $\lim_{\v{q}\to 0} \lim_{\Omega_m\to 0} Q^{ij}_{\mathrm{fluc}}(q) = 0$ respectively.
    Adding additional diagrams to this response has to be done with caution not to conflict with these physical constraints.

    In momentum space, the fluctuation response kernel $Q^{\mu\nu}_{\mathrm{fluc}}(q) = \int \D (x-x') \e^{-\iu q \cdot (x-x')} Q_{\mathrm{fluc}}^{\mu\nu}(x,x')$ is decomposed as
    $
        Q^{\mu\nu}_{\mathrm{fluc}}(q) = Q^{\mu\nu}_{\mathrm{AL}}(q) + Q^{\mu\nu}_{\mathrm{DOS}}(q) +Q^{\mu\nu}_{\mathrm{MT}}(q) + Q^{\mu\nu}_{\mathrm{DIA}}(q),
    $
    where
    \begin{widetext}
        \begin{subequations}
            \begin{align}
                 \begin{split}
                    Q^{\mu\nu}_{\mathrm{AL}}(q)
                    &= - \frac{e_{*}^2 }{64} \frac{1}{(\beta V)^3} \sum_{p k k'} D(p) D(p-q) G_{0}(k) \gamma^{\mu}(k,k-q) G_{0}(k-q) \left[ G_{0}(k+p-q) - G_{0}(k-p) \right] \\
                    & \hspace{13em}\times G_{0}(k') \gamma^{\nu}(k',k'-q) G_{0}(k'-q) \left[ G_{0}(k'+p-q) - G_{0}(k'-p) \right] 
                 \end{split}\\
                 \begin{split}
                     Q^{\mu\nu}_{\mathrm{DOS}}(q) &= - \frac{e_{*}^2}{8} \frac{1}{(\beta V)^2} \sum_{kp} D(p) \left[G_{0}(k)\right]^2 G_{0}(k+p) \Big[ \gamma^{\mu}(k,k+q) G_{0}(k+q) \gamma^{\nu}(k+q,k) \\
                     &\hspace{17.6em} +  \gamma^{\mu}(k,k-q) G_{0}(k-q) \gamma^{\nu}(k-q,k) \Big]
                 \end{split} \\
                      Q^{\mu\nu}_{\mathrm{MT}}(q) &= \frac{e_{*}^2}{8} \frac{1}{(\beta V)^2} \sum_{kp} D(p) G_{0}(k-q) \gamma^{\mu}(k-q,k) G_{0}(k) G_{0}(k+p) \gamma^{\nu}(k+p,k+p-q) G_{0}(k+p-q) \\
                 Q^{\mu\nu}_{\mathrm{DIA}}(q) &= - \frac{e_{*}^2}{8 m} \delta^{\mu\nu} \left( 1 - \delta^{\nu 0} \right) \frac{1}{(\beta V)^2} \sum_{kp} D(p) \left[ G_{0}(k) \right]^2 G_{0}(k+p).
            \end{align}
        \end{subequations}
    \end{widetext}
    Here $k$ and $k'$ are four-momenta with Fermionic Matsubara frequencies, while $p$ is one with a Bosonic Matsubara frequency. The summation involves a sum over the Matsubara frequency and the momentum $\sum_{k} \coloneqq \sum_{n\in\ZZ} \sum_{\v{k}}$.
    Moreover, we have introduced the Fermionic current vertex $\gamma^{\mu}(k,k-q) = (1, \frac{1}{m}(\v{k} - \v{q}/2))$ and denoted the charge of the electrons under the spin gauge field by $e_{*}$. 
    
    \paragraph{Fluctuation spin conductivity.}\label{sec:sigma}

    A spin current can be generated by a gradient of a magnetic field 
    \begin{equation}
        J^{i}_{z}(\omega) = \sigma^{ij}_{z}(\omega) \left( \partial_{j} B^z \right)(\omega).
    \end{equation}
    Noting that the temporal component of the gauge potential couples to the electron spin density as $-B^{z}$ \cite{Tokatly-Tokatly-EquilibriumSpinCurrents-2008}, we can obtain the spin conductivity from the response kernel as $\sigma^{ij}_{z}(\omega) = \lim_{\v{q}\to0} \iu \, \partial Q^{i0}(q)/\partial q^{j}$.
    However, the gauge invariance of the current in Eq.~\eqref{eq:J_general} dictates that $A_{0}$ only enters through the combination $F_{0i} = \partial_{0} A_{i} - \partial_{i} A_{0}$.
    This can be leveraged to access a simpler way of computing $\sigma^{ij}_{z}(\omega)$ which parallels the computation of electrical conductivity in the Weyl gauge \cite{Altland-Simons-CondensedMatterField-2023}
    \begin{equation}\label{eq:sigma_relation}
        \sigma^{ij}_{z}(\omega) = \lim_{\v{q}\to 0}\frac{\iu}{\omega + \iu 0} Q^{ij}(\omega,\v{q}).
    \end{equation}
    We stress that using the simplifications afforded by gauge invariance is only permitted once the response tensor is shown to satisfy $q_{\mu} Q^{\mu\nu}(q) = 0$.

    The fluctuation corrections to the electrical conductivity at finite frequency of a single-band metal sum to zero in the absence of disorder \cite{Boyack-Boyack-RestoringGaugeInvariance-2018,Lundemo-Sudbo-FluctuationConductivityUltraclean-2026b}.
    In the Supplemental material we demonstrate this explicitly in the present context \cite{--SeeSupplementalMaterial-}.
    This symmetry protection follows from the fact that the uniform electrical current is proportional to the total momentum, which does not relax in a Galilean-invariant system \cite{Gindikin-Maslov-QuantumCriticalityOptical-2024}.
    In a multiband metal, however, this constraint is lifted and a finite dissipative part of the conductivity is allowed, even in an ultraclean system \cite{Lundemo-Sudbo-FluctuationConductivityUltraclean-2026b}. 
    The close analogy to the fluctuation spin conductivity implies that interactions involving momentum transfer between spin up and down electrons provide a mechanism for spin current relaxation.
    The real part of Eq.~\eqref{eq:sigma_relation} is therefore generically nonzero at finite frequencies.
    
    By studying the fluctuation diagrams in the uniform limit, we find that the MT and DOS diagrams partially cancel each other and leave a term that can be brought to a form similar to the AL diagrams \cite{--SeeSupplementalMaterial-}.
    In contrast to the case of the fluctuation conductivity of a superconductor, there is not a direct relation between the leftover term from the sum of MT and DOS and the AL diagrams, which obstructs a cancellation of the type discovered in the former case \cite{Lundemo-Sudbo-FluctuationConductivityUltraclean-2026b}.
    In the superconducting fluctuation theory, it is commonly believed that the AL diagrams dominate the response kernel in the critical regime, being constructed from two fluctuation propagators with a pole at the critical point, as opposed to one.
    This rationale was recently shown to fail in an ultraclean metal, where all the Gaussian-level fluctuation diagrams contributing to the conductivity were shown to be of the same order \cite{Lundemo-Sudbo-FluctuationConductivityUltraclean-2026b,Boyack-Boyack-RestoringGaugeInvariance-2018}.
    As for the spin conductivity, a similar conclusion arises: all the fluctuation diagrams arising from the Gaussian-fluctuation approximation give equally singular contributions to the spin conductivity in the critical regime.
    A similar conclusion was recently reached in a calculation of the particle-hole susceptibilities of a metal approaching a Pomeranchuk instability \cite{Gindikin-Chubukov-CollectiveExcitationsStability-2025}.  

    After adding all the Gaussian-fluctuation diagrams together we can transform the Matsubara summation to a contour integral in the standard fashion \cite{--SeeSupplementalMaterial-,Altland-Simons-CondensedMatterField-2023}.
    The real part of the longitudinal spin-$z$ conductivity at finite frequencies consists of two parts
    \begin{equation}\label{eq:conductivity_sum}
        \Re \sigma_{z}(\omega) = \Re \sigma_{z}^{\mathrm{qp}}(\omega) + \Re \sigma_{z}^{\mathrm{fluc}}(\omega),
    \end{equation}
    where $\Re \sigma_{z}^{\mathrm{qp}}(\omega)$ is a contribution to the quasiparticle transport derived from parts of the MT and DOS diagrams while $\Re \sigma_{z}^{\mathrm{fluc}}(\omega)$ receives contributions from all diagrams and represents a new channel of spin transport constituted by the spin fluctuations themselves.
    The Gaussian-fluctuation approximation of a clean system captures only the $\omega \gg \Gamma $ tail of the former \cite{Kim-Lee-GaugeinvariantResponseFunctions-1994}
    \begin{equation}
        \Re \sigma_{z}^{\mathrm{qp}}(\omega) \propto \frac{\Gamma}{\Gamma^2 + \omega^2} \approx \frac{\Gamma}{\omega^2}.
    \end{equation}
    In the limit $\alpha \xi^{-3} \ll 1 $, where $\alpha = 2 \epsilon_F/(\gamma T)$ and $\gamma$ is the coefficient of the Landau damping, we find that $\Gamma \propto \alpha^{-1} \log(\xi^{3}/\alpha )$, which near the quantum critical point implies $\Gamma \sim T \log(\epsilon_F/T)$.
    In the opposite limit $\alpha \xi^{-3} \gg 1$ we find $\Gamma \propto T^2 \xi^3$ \cite{--SeeSupplementalMaterial-}. 

    The fluctuation spin conductivity is given by 
    \begin{align}\label{eq:sigma_integral}
        \Re \sigma_{z}^{\mathrm{fluc}}(\omega) \approx \frac{C e_{*}^2}{\omega} &\int \frac{\D^3 p}{(2\pi)^3} \int \frac{\D z}{2\pi} V_1^{R}(z,\v{p}) V_2^{R}(z,\v{p}) \notag \\
        \times & \left[ \coth\left( \frac{z}{2 T}\right) - \coth\left( \frac{z-\omega}{2 T}\right)  \right] \notag \\
        \times & \Im D^{R}(z,\v{p}) \Im D^{R}(z-\omega,\v{p}),
    \end{align}
    where $C$ is a numerical prefactor.
    In the critical regime, the product of the vertices here has a leading contribution $V_1^{R}(z,\v{p}) V_{2}^{R}(z,\v{p}) \sim -1/p^2$ while the propagators are defined by Eq.~\eqref{eq:D_propagator_inverse}.
    In general, the fluctuation spin conductivity can be expressed in terms of a scaling function of two variables $\mathcal{F}(x,y)$ as
    \begin{equation}
        \Re \sigma_z^{\mathrm{fluc}}(\omega) \propto T \xi^{6} \mathcal{F}(\tilde{\omega}\xi^{3}, \alpha \xi^{-3}),
    \end{equation}
    where $\tilde{\omega} = \gamma \omega/(4\epsilon_F)$.
    In the quantum critical (QC) regime of Fig.~\ref{fig:schematic_pd} we can take the limit $\alpha \xi^{-3} \ll 1$ where the spin-$z$ conductivity can be expressed in terms of a scaling function $\Phi(x)$ as
    \begin{equation}\label{eq:Re_sigma_scaling_form}
        \Re \sigma_z^{\mathrm{fluc}}(\omega) \propto T \xi^{6} \Phi(\tilde{\omega}\xi^{3}).
    \end{equation}
    This function is shown in Fig.~\ref{fig:scaling_function} and derived in the Supplemental material \cite{--SeeSupplementalMaterial-}.
    Notably, in the low-frequency limit $\tilde{\omega}\xi^{3} \ll 1$ the most singular contribution is given by
    \begin{equation}\label{eq:sigma_low_freq_QC}
        \Re \sigma_z^{\mathrm{fluc}}(\omega) \sim T \xi^{6} \log\left(\frac{1}{\tilde{\omega}\xi^{3}}\right).
    \end{equation}

    Well below the crossover line $T \sim (\delta g)^{3/2}$ in Fig.~\ref{fig:schematic_pd} we have $\alpha \xi^{-3} \gg 1$, which corresponds to the Fermi-liquid regime \cite{Li-Maslov-OpticalConductivityMetal-2023}. 
    In the low-frequency limit $\tilde{\omega} \xi^{3} \ll 1$ the most singular contribution now takes a similar form
    \begin{equation}\label{eq:sigma_low_freq_FL}
        \Re \sigma_z^{\mathrm{fluc}}(\omega) \sim T \xi^{6} \log\left(\frac{T}{\omega}\right).
    \end{equation}
    Equations \eqref{eq:sigma_low_freq_QC} and \eqref{eq:sigma_low_freq_FL} are the main results of this paper.
    As $\xi\to \infty$, the spin conductivity receives a critical enhancement in a narrow frequency range around $\omega=0$, ultimately suggesting a divergent dc limit upon approaching the QCP. 
    The logarithmic divergence as $\omega \to 0$ derives from the infinite lifetime of the Landau-damped spin fluctuations in the infrared limit and is therefore expected to be regularized once effects providing a finite lifetime of these modes are included in the model.    
    
    \begin{figure}[t]
        \centering
        \includegraphics[width=\linewidth]{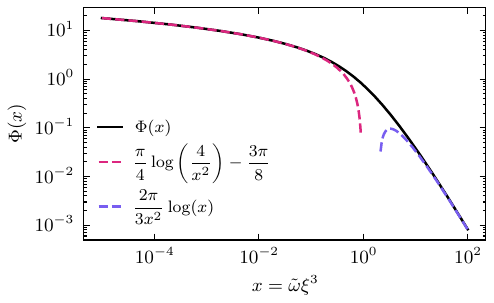}
        \caption{Scaling function $\Phi(x)$ that determines the frequency dependence of the conductivity in the limit $\alpha\xi^{-3}\ll1$, computed from Eq.~\eqref{eq:sigma_integral}.
        The figure also includes the asymptotic behaviors of $\Phi(x)$ as $x\ll 1$ and $x\gg1$.
        }
        \label{fig:scaling_function}
    \end{figure}
    
    While similar critical enhancement effects have previously been found in classical or impure local-moment systems \cite{Aoyama-Aoyama-SpinThermalTransport-2022,Okamoto-Nagaosa-CriticalSpinFluctuation-2019,Okamoto-Nagaosa-CriticalEnhancementSpin-2024}, we show this in an interacting fermionic theory with neither disorder nor spin-orbit coupling. 
    Importantly, this demonstrates that the mechanism for the spin conductivity enhancement is purely intrinsic.
    The added appeal of considering a clean system is that it allows for a controlled calculation of the spin conductivity in the Gaussian-fluctuation approximation, in the sense that it represents a conserving approximation. 
    Going beyond the Gaussian-fluctuation approximation, it is well known that non-analytic corrections from soft modes may render the transition first-order at the lowest temperatures \cite{Belitz-Kirkpatrick-QuantumCriticalBehaviour-1996,Belitz-Vojta-FirstOrderTransitions-1999,Chubukov-Rech-InstabilityQuantumCriticalPoint-2004,Rech-Chubukov-QuantumCriticalBehavior-2006}.
    We expect our results to apply above this scale, where the Hertz-Millis framework is applicable. 
    Moreover, we do not expect the effects of vertex corrections and self energies beyond this level of approximation to change the results qualitatively.
    Indeed, such effects were recently shown to cancel one another in a related study of particle-hole susceptibilities in the charge channel \cite{Gindikin-Chubukov-CollectiveExcitationsStability-2025}.
    
    This critical enhancement of the spin conductivity should be contrasted with the results of Ref.~\cite{Duine-Vignale-SpinDragUltracold-2010} which suggest that the spin conductivity is instead \textit{reduced} upon approaching a classical ferromagnetic transition.
    That calculation was based on a semiclassical Boltzmann approach to the calculation of the spin-drag relaxation rate $\Gamma_{\mathrm{sd}}$, which was found to diverge close to the classical transition. 
    Applied to the QCP, however, their relaxation rate also decreases as $T\to 0$.
    That calculation treats spin fluctuations only as a scattering mechanism for electrons and is by construction oblivious to the processes captured by the second term of Eq.~\eqref{eq:conductivity_sum} that describe the contribution of the $\mathrm{U}(1)_z$-charged collective excitations to the spin conductivity. 
    Moreover, Ref.~\cite{Duine-Vignale-SpinDragUltracold-2010} considers an $\mathrm{SU}(2)$-symmetric model and treats all scattering processes as spin-conserving, giving rise to a weight $\sim p^2$ in the integrand.
    In contrast, the transverse fluctuations we consider are spin-flip processes and carry no such suppression.
    This is reflected by a delicate cancellation between the MT and DOS diagrams in our case \cite{--SeeSupplementalMaterial-}.

    That the fluctuation spin conductivity receives divergent corrections close to the QCP is interpreted as a precursor to the spin superfluidity of the ordered state.
    The classical picture of spin superfluidity in easy-plane ferromagnets can be understood in terms of the current-loop representation of the XY model
    \begin{align}
        Z_{XY} &= \int \DD \theta \sum_{\{S\}} \e^{- \frac{\beta}{2} \int \D^3 x \, \left( \bm{\nabla}  \theta(x) - 2\pi \bm{\delta}(x;S) \right)^2  } \notag \\
        &\sim \sum_{\{L \, \mathrm{closed} \}} \e^{- \frac{1}{2\beta} \int \D^3 x \, \v{b}^2(x) },
    \end{align}
    where the compactness of the phase $\theta$ is accounted for by including delta functions on surfaces $S$ across which the phase jumps by $2\pi$ and $\v{b}(x) = \bm{\delta}(x;L)$ is a delta function on the boundary $L$ of these surfaces \cite{Kleinert-Kleinert-MultivaluedFieldsCondensed-2008}.
    When approaching the ordering temperature from above, larger current loops $L$ receive a significant Boltzmann weight, eventually extending across the entire system in the superfluid phase.
    Although an exact loop-current mapping is not possible in our case, we argue that the same intuitive picture may apply; in our case the statistical weight of spin-current loops in the quantum critical region is enhanced by dynamical order-parameter fluctuations with dynamical critical exponent $z=3$ born out of the Fermi sea. 
    
    The analogous phenomenon in superconductors, called \textit{paraconductivity}, refers to the excess conductivity gained by the opening of a new channel of charge transport by fluctuating Cooper pairs above $T_c$ \cite{Larkin-Varlamov-TheoryFluctuationsSuperconductors-2005}. 
    Here, the spin fluctuations in the normal state play a role similar to the fluctuating Cooper pairs, since they are charged under the spin-rotational $\mathrm{U}(1)_z$ symmetry.
    A crucial difference between the two phenomena is that they belong to different dynamical universality classes.
    Indeed, the classical superconducting fluctuations are characterized by a dynamical exponent of $z=2$ \cite{Larkin-Varlamov-TheoryFluctuationsSuperconductors-2005}, while $z=3$ for the ferromagnetic quantum spin fluctuations.
    This affects the scaling of the fluctuation corrections to the conductivities. 
    Fundamentally, a direct analogy between the two phenomena is moreover spoiled by the fact that a superconducting QCP is rather different from a ferromagnetic one: driving $T_c\to 0$ in a pure BCS superconductor eliminates the pairing interaction as well as the fluctuations altogether \cite{Ramazashvili-Coleman-SuperconductingQuantumCritical-1997}.  
    To obtain a superconducting QCP, one generically has to rely on pair-breaking disorder \cite{Ramazashvili-Coleman-SuperconductingQuantumCritical-1997,Herbut-Herbut-ZeroTemperature$mathitd$WaveSuperconducting-2000,Mineev-Sigrist-CriticalFluctuationEffects-2001,Dzero-Levchenko-TransportAnomaliesMultiband-2023}.

    \paragraph{Concluding remarks.}\label{sec:conclusion}

    By employing the simplest microscopic model of interacting electrons that supports a ferromagnetic QCP \cite{Stoner-Stoner-CollectiveElectronFerromagnetism-1938}, we have demonstrated that the spin conductivity receives a critical enhancement as the QCP is approached. 
    We demonstrated the soundness of our theory by proving that the fluctuation corrections obey the Ward identity that derives from the continuity equation for the spin density, and that the spin stiffness vanishes in the normal state.
    The critical enhancement of the conductivity close to the QCP is indicative of the proximity to a phase supporting coherent spin transport.  
    This is interpreted as the spin analog of the paraconductivity phenomenon in superconductors \cite{Aslamazov-Larkin-EffectFluctuationsProperties-1968,Aslamazov-Larkin-InfluenceFluctuationPairing-1968,Maki-Maki-CriticalFluctuationOrder-1968,Thompson-Thompson-MicrowaveFluxFlow-1970}.
    This work therefore establishes a new link between spin-transport phenomena and the theory of superconducting fluctuations, which might prove useful for probing the precursor to quantum critical behavior in itinerant magnets.
    Importantly, it also sheds new light on the phenomenon of spin superfluidity by bringing it into the quantum regime.
    
	\begin{acknowledgments}
		We acknowledge support from the Norwegian Research Council through Grant No. 262633, ``Center of Excellence on Quantum Spintronics” and Grant No. 323766, as well as COST Action CA21144 ``Superconducting Nanodevices and Quantum Materials for Coherent Manipulation".
	\end{acknowledgments}
	
	\bibliography{spin_conductivity_references}

    \clearpage
    \onecolumngrid

\renewcommand{\thefigure}{S\arabic{figure}}
\renewcommand{\theHfigure}{S\arabic{figure}}
\setcounter{figure}{0}  

\renewcommand{\thetable}{S\Roman{table}}
\setcounter{table}{0}  

\renewcommand{\theequation}{S\arabic{equation}}
\setcounter{equation}{0}  

\setcounter{secnumdepth}{3}

\renewcommand{\thesection}{S\arabic{section}}
\renewcommand{\thesubsection}{\thesection.\arabic{subsection}}
\renewcommand{\thesubsubsection}{\thesubsection.\arabic{subsubsection}}

\makeatletter
\renewcommand{\p@subsection}{}
\renewcommand{\p@subsubsection}{}
\makeatother

% \title{
%     Supplemental material for \\
%     ``Divergent spin conductivity on the verge of ferromagnetic quantum criticality" 
%     }
	
% 	\author{Sondre Duna Lundemo} 
% 	\affiliation{Center for Quantum Spintronics, Department of Physics, Norwegian University of Science and Technology, NO-7491 Trondheim, Norway}
	
% 	\author{Asle Sudb\o}
% 	\affiliation{Center for Quantum Spintronics, Department of Physics, Norwegian University of Science and Technology, NO-7491 Trondheim, Norway}
	
% 	\date{\today} 

% \begin{abstract}
%     This supplemental material provides details on the construction of the Gaussian-fluctuation action, the resulting fluctuation response kernel, as well as a detailed calculation of the fluctuation spin conductivity that derives from it.
% \end{abstract}
% \maketitle
% \onecolumngrid
% \vspace{-3em}
% \begingroup
% \renewcommand{\cftsecfont}{\normalfont\bfseries}
% \renewcommand{\cftsecpagefont}{\normalfont\bfseries}
% \hypersetup{linkcolor=black}
% \tableofcontents
% \endgroup

%--------------------------------------------------------------------------------------------------
%only for arxiv version:
\begin{center}
	{
		{\large \bfseries
			Supplemental material for \\
			``Divergent spin conductivity on the verge of ferromagnetic quantum criticality" }
	}    
	\vspace{1em}
	
	Sondre Duna Lundemo and Asle Sudb\o 
	
	\textit{Center for Quantum Spintronics, Department of Physics, \\ Norwegian University of Science and Technology, NO-7491 Trondheim, Norway}
	
	(Dated: \today)
	
\end{center}

\begin{adjustwidth}{3.5em}{3.5em}
	\quad This supplemental material provides details on the construction of the Gaussian-fluctuation action, the resulting fluctuation response kernel, as well as a detailed calculation of the fluctuation spin conductivity that derives from it.
\end{adjustwidth}

\begingroup
\renewcommand{\cftsecfont}{\normalfont\bfseries}
\renewcommand{\cftsecpagefont}{\normalfont\bfseries}
\hypersetup{linkcolor=black}
\startcontents[rest]                 
\printcontents[rest]{ }{2}{}         
\endgroup
%--------------------------------------------------------------------------------------------------

\section{Fluctuation action and propagator}\label{suppl:fluct}

In this section, we derive the fluctuation propagator and the fluctuation action in the Gaussian-fluctuation approximation.

Consider an itinerant magnet described by the fermionic coherent state functional integral in the presence of an external (non-dynamical) spin-$z$ gauge field $A_{\mu}$ 
\begin{equation}
	Z[A] = \int \DD \bar{\psi} \DD \psi \e^{-S[\bar{\psi},\psi,A]},
\end{equation}
where the action takes the form
\begin{equation}
	\begin{split}
		S[\bar{\psi},\psi,A] &= \int \D x \int \D y \sum_{\sigma} \bar{\psi}_{\sigma}(x) \left(- G^{-1}_{0}[A](x,y)\right) \psi_{\sigma}(y) \\
		&- \int \D x \, \left[ \frac{J}{2} \Big( S_{x}(x) S_{x}(x) + S_{y}(x) S_{y}(x)  \Big) - K_{z}   S_{z}(x) S_{z}(x) \right].
	\end{split}
\end{equation}
The interaction comprises an isotropic, ferromagnetic XY interaction between the electron spins $S_{\alpha}(x) \coloneqq (\psi^{\dagger} \sigma^{\alpha} \psi)(x) /2$ and an easy-plane anisotropy $K_z >0$.
Here, we have introduced the Grassmann spinor $\psi(x) \coloneqq \left( \psi_{\uparrow}(x) \quad \psi_{\downarrow}(x) \right)^{\mathsf{T}}$ and the bare Green's function (with $A = 0$)
\begin{equation}
	G_{0}^{-1}(x,y) = - \left[ \partial_{\tau} + \xi(-\iu\bm{\nabla}) \right] \delta(x-y).
\end{equation}
The easy-plane anisotropy $K_z > 0$ ensures that the saddle point of the electron magnetization is characterized by $\langle S_{z} \rangle = 0$. 
Moreover, the $S_z$ fluctuations about this state are gapped and can therefore be omitted in the following.

The bi-quadratic XY interaction is now eliminated in favor of a linear coupling between the electron spin density and an auxiliary boson through a Hubbard-Stratonovich decoupling in the spin-density channel.
This is achieved by introducing the two-component real field $\bm{\pi} \coloneqq (\pi_x, \pi_y)$ together with a functional-integral measure normalized so that
\begin{equation}
	1 = \int \DD[ \bm{\pi}] \exp \left( - \int \D x \frac{\bm{\pi}^2(x)}{2 J} \right).
\end{equation}
Inserting this into the partition function and performing the shift $
	\pi_{i}(x) \mapsto \pi_{i}(x) - J S_{i}(x)
$ yields the new action
\begin{equation}
	S_{\mathrm{HS}}[\bar{\psi},\psi,\bm{\pi},A] = \int \D x \int \D y \, \psi^{\dagger}(x) \Big(- \mathcal{G}^{-1}_{0}[A](x,y) + \Sigma[\bm{\pi}](x,y) \Big) \psi(y) + \int \D x \frac{\bm{\pi}^2(x)}{2 J}.
\end{equation}
In this equation, we have denoted $\mathcal{G}^{-1}_{0}[A] =  G_{0}^{-1}[A] \bm{1}$ and $\Sigma[\bm{\pi}](x,y)  = -\delta(x-y) \bm{\pi}(x) \cdot \bm{\sigma} /2 = -\delta(x-y)\left( \bar{\pi}(x) \sigma^{+} + \pi(x) \sigma^{-}  \right)/2$ with the two-by-two raising and lowering matrices $\sigma^{\pm} \coloneqq (\sigma^{1} \pm \iu \sigma^{2})/2$.
The complex-field parametrization of the auxiliary field is related to the real one by $\pi(x) = \pi_{x}(x) + \iu \pi_{y}(x)$ and $\bar{\pi}(x) = \pi_{x}(x) - \iu \pi_{y}(x)$.
This parametrization is particularly convenient to see the analogy with the present problem and that of a fermionic superfluid \cite{Lundemo-Sudbo-FluctuationConductivityUltraclean-2026b,Boyack-Boyack-RestoringGaugeInvariance-2018}.
After performing the integral over the fermionic fields we find the partition function
\begin{equation}
	Z[A] = \int \DD [\bm{\pi}] \exp\left( -S_{\mathrm{eff}}[\bm{\pi},A] \right),
\end{equation}
where 
\begin{equation}
	S_{\mathrm{eff}}[\bm{\pi},A] =  - \tr \log \left( -\beta \mathcal{G}^{-1}[A, \bm{\pi}] \right) + \int \D x \frac{\bm{\pi}^2(x)}{2 J},
\end{equation}
and $\mathcal{G}^{-1}[A,\bm{\pi}](x,y) \coloneqq \mathcal{G}_{0}^{-1}[A](x,y) - \Sigma[\bm{\pi}](x,y)$.

\subsection{Saddle point}

Before we can proceed with the Gaussian-fluctuation expansion, we need to identify the correct saddle point for $A \neq 0$.
In general, the magnetization of a Fermi liquid is a functional of the external spin gauge field which cannot be put to $0$ in this calculation, even if the analysis is performed above $T_c$.
This is because we are ultimately interested in the linear response to the external gauge potential and the implicit dependence on $A$ via the mean-field magnetization introduces new collective mode terms in the response kernel.
This is analogous to what happens in a superconductor below the ordering temperature or in nonuniform superconductors, see Refs.~\cite{Anderson-Levin-GoingBCSLevel-2016,Boyack-Levin-CollectiveModeContributions-2017,Millis-Millis-MeissnerEffectAnisotropic-1987} for a detailed account of this.
In the easy-plane limit, however, the expansion around $\bm{\pi}_{\mathrm{MF}} = 0$ above $T_c$ is permitted since the response function that enters these collective mode terms is given by
\begin{equation}
	\frac{\delta \bm{\pi}_{\mathrm{MF}}[A](x)}{\delta A_{\mu}(y)} \biggr\lvert_{A = 0} = - 2 J \tr \left( \mathcal{G}[0,\bm{\pi}] \frac{\delta \mathcal{G}^{-1}[A,\bm{\pi}]}{\delta A_{\mu}(y)} \mathcal{G}[0,\bm{\pi}] \frac{\delta \mathcal{G}^{-1}[A,\bm{\pi}]}{\delta \bm{\pi}(x) } \right) \biggr\lvert_{\bm{\pi} = \bm{\pi}_{\mathrm{MF}}[A],A = 0} = 0.
\end{equation}
Conversely, considering the full $\mathrm{SU}(2)$-invariant exchange interaction would require taking these collective-mode terms into account to obtain a gauge-invariant response function \cite{Anderson-Levin-GoingBCSLevel-2016}.

\subsection{Gaussian fluctuation action}

Above the ordering temperature, we parametrize the magnetization as $\bm{\pi}(x) = \bm{\pi}_{\mathrm{MF}}(x) + \bm{\phi}(x) \equiv \bm{\phi}(x)$ and expand the action to quadratic order in $\bm{\phi}$.
This yields
\begin{equation}
	S_{\mathrm{eff}}[\bm{\phi}, A] = - \tr \log\left(-\beta \mathcal{G}^{-1}_{0}[A]\right) + \int \D x \int \D y \, \bar{\phi}(x) \left[ - D^{-1}[A](x,y) \right] \phi(y),
\end{equation}
where 
\begin{align}
	 - D^{-1}[A](x,y)  &= \frac{\delta(x-y)}{2 J} - \frac{\delta^2}{\delta \bar{\pi}(x) \delta \pi(y)} \tr\log\left(-\beta \mathcal{G}^{-1}[A,\bm{\pi}]\right) \biggr\lvert_{\bm{\pi} = 0} \notag \\
	&=\frac{\delta(x-y)}{2 J} - \int \D z \int \D z' \tr \left( \frac{\delta \mathcal{G}[A,\bm{\pi}](z,z')}{\delta \bar{\pi}(x)} \frac{\delta \mathcal{G}^{-1}[A,\bm{\pi}](z',z)}{\delta \pi(y)} \right)\biggr\lvert_{\bm{\pi} = 0}.
\end{align}
Using the identity $\int \D x' \mathcal{G}(x,x') \mathcal{G}^{-1}(x',y) = \delta(x-y)$ we find
\begin{align}
	- D^{-1}[A](x,y) &= \frac{\delta(x-y)}{2 J} \notag\\
	&+ \int \prod_{i=1}^{4} \D z_{i} \tr \left( \mathcal{G}[A,0](z_1,z_2) \frac{\delta \mathcal{G}^{-1}[A,\bm{\pi}](z_2,z_3)}{\delta \bar{\pi}(x)} \mathcal{G}[A,0](z_3,z_4) \frac{\delta \mathcal{G}^{-1}[A,\bm{\pi}](z_4,z_1)}{\delta \pi(y)} \right)\biggr\lvert_{\bm{\pi} = 0}.
\end{align}
Performing the functional derivatives now yields 
\begin{equation}\label{eq:D_traceformula}
		- D^{-1}[A](x,y) = \frac{\delta(x-y)}{2 J} + \frac{1}{4} \tr \left( \mathcal{G}[A,0](y,x) \sigma^{+} \mathcal{G}[A,0](x,y) \sigma^{-}  \right).
\end{equation}
The form of the propagator shown in Eq.~\eqref{eq:D_traceformula} will be useful later in Sec.~\ref{suppl:response}.
After integrating out the bosonic fields we are left with the effective action
\begin{equation}
    S_{\mathrm{eff}}[A] = - \tr \log\left(-\beta \mathcal{G}^{-1}_{0}[A]\right) + \tr \log\left(-\beta D^{-1}[A]\right).
\end{equation}

\subsection{Fluctuation propagator}

Let us now put $A = 0$ and consider the fluctuation propagator in more detail.
Performing the trace yields
\begin{equation}
	D^{-1}(x,y) = - \frac{\delta(x-y)}{2 J} - \frac{1}{4} G_{0}(x,y) G_{0}(y,x).
\end{equation}
To write the action in the momentum space, we introduce the Fourier transform of the field $\phi$, and its inverse as
\begin{equation}
    \phi(x) = \frac{1}{\sqrt{\beta V}} \sum_{q} \e^{\iu q \cdot x} \phi(q) \quad \text{and} \quad \phi(q) = \frac{1}{\sqrt{\beta V}} \int \D x \, \e^{-\iu q \cdot x} \phi(x).
\end{equation}
We have defined the four-momentum as $q^{\mu} \equiv (q_0, \v{q}) = ( \iu \Omega_m, \v{q})$ where $\Omega_m$ is a bosonic Matsubara frequency and the summation is $\sum_{q} \equiv \sum_{m \in \ZZ } \sum_{\v{q}}$.
Note that $q \cdot x \coloneqq \v{q} \cdot \v{r} - \Omega_m \tau \equiv - \mathfrak{g}_{\mu\nu} q^{\mu} x^{\nu}$.
With these definitions at hand, we find
\begin{equation}
    S_{\mathrm{eff}}[\bm{\phi},A ] = - \tr \log\left(-\beta \mathcal{G}^{-1}_{0}[A]\right) + \sum_{q} \bar{\phi}(q) \left[ - D^{-1}(q) \right] \phi(q),
\end{equation}
with
\begin{align}\label{eq:Dinv}
    D^{-1}(q) = \int \D (x-y) \, \e^{-\iu q \cdot (x-y)} D^{-1}(x-y) = -\frac{1}{2J} \left[ 1 + \frac{J}{2} \frac{1}{\beta V}\sum_{k} G_{0}(k) G_{0}(k-q) \right] \equiv - \frac{1}{2J} \left[ 1 + \frac{J}{2} \Pi(q) \right],
\end{align}
where we have introduced the particle-hole bubble $\Pi(q)$ in the last transition. 
The Fourier transform of the Green's function is likewise defined by
\begin{equation}
    G_{0}(x,x') = \frac{1}{\beta V} \sum_{k} \e^{\iu k \cdot (x-x')} G_{0}(k).
\end{equation}
The Gaussian-fluctuation approximation reproduces the ladder approximation of the Dyson equation for the propagator of the spin fluctuations, which can be seen by inverting Eq.~\eqref{eq:Dinv}
\begin{equation}
    \frac{D(q)}{2 J} = - \left[ 1 + \frac{J}{2} \Pi(q) \frac{D(q)}{2J} \right].
\end{equation}

\subsection{Particle-hole bubbles}\label{suppl:ph_bubbles}

\begin{figure*}[htb]
    \centering
    \begin{minipage}{0.5\linewidth}
        \centering
        \includegraphics[width=\linewidth]{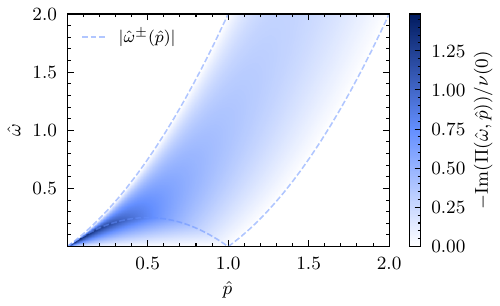}
    \end{minipage}~
    \begin{minipage}{0.5\linewidth}
        \centering
        \includegraphics[width=\linewidth]{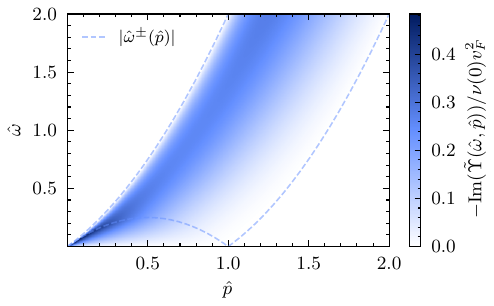}
    \end{minipage}
    \caption{Imaginary part of the particle hole bubbles $\Pi(\omega,p)$ (left-hand panel) and $\tilde{\Upsilon}(p)$ (right-hand panel) as a function of the dimensionless frequency $\hat{\omega} \equiv \omega/4 \epsilon_F$ and dimensionless momentum $\hat{p} \equiv p/2 k_F$. The dashed lines show the curves $\hat{\omega}^{\pm}(\hat{p}) = \pm \abs{\hat{p}} + \hat{p}^2$ that define the borders of the particle-hole continuum.}
    \label{fig:ph_bubbles}
\end{figure*}

The particle-hole bubble that enters in the spin-fluctuation propagator is the familiar Lindhard function
\begin{align}\label{eq:ph_bubble}
    \Pi(\iu\Omega_m, \v{p}) &= T \sum_{n \in \ZZ} \int \frac{\D^3 k}{(2\pi)^3} G_{0}(k) G_{0}(k-p) = \int \frac{\D^3 k}{(2\pi)^3} \frac{n(\v{k}-\v{p}) - n(\v{k})}{\xi(\v{k}-\v{p}) - \xi(\v{k}) + \iu \Omega_m},
\end{align}
where $n(\v{k})\equiv n(\xi(\v{k}))$ denotes the Fermi distribution function.
At $T = 0$, this can be computed exactly as ($\hat{p} \coloneqq p /2k_F, \hat{\omega} \coloneqq \omega/4 \epsilon_F$)
\begin{align}\label{eq:Pi(p)}
    \Pi(\omega, \v{p}) &= \frac{1}{4\pi^2} \int_{0}^{k_F} \D k \, k^2 \int_{-1}^{+1} \D \cos(\theta) \frac{1}{kp \cos(\theta)/m - p^2 /(2m) - \omega} + (p \leftrightarrow -p, \omega \leftrightarrow -\omega)\notag \\
    &= - \frac{1}{4 \hat{p}} \frac{m k_F}{2\pi^2} \int_{0}^{1} \D x \, x^2 \int_{-1}^{+1} \D c \frac{1}{\hat{p} + \hat{\omega}/\hat{p} - x c } + \left( \hat{p} \leftrightarrow - \hat{p}, \hat{\omega} \leftrightarrow - \hat{\omega} \right) \notag \\
    &= -\frac{\nu(0)}{2} -  \frac{\nu(0)}{2} \left\{\frac{1}{4 \hat{p}} \left[ 1 - \left( \hat{p} + \frac{\hat{\omega}}{\hat{p}} \right)^2 \right] \log\left(\frac{\hat{p} + \hat{\omega}/\hat{p} + 1}{\hat{p} + \hat{\omega}/\hat{p} - 1}\right) + \left( \hat{p} \leftrightarrow - \hat{p}, \hat{\omega} \leftrightarrow - \hat{\omega} \right) \right\}.
\end{align}
Here $\nu(0)$ denotes the density of states at the Fermi level.
The imaginary part of $\Pi(\omega + \iu 0, \v{p})$ is shown on the left-hand side of Fig.~\ref{fig:ph_bubbles}, together with the boundaries of the particle-hole continuum $\hat{\omega}^{\pm}(\hat{p}) = \pm \abs{\hat{p}} + \hat{p}^2$.

In the fluctuation response kernel, we will also encounter the particle-hole bubble with velocity insertions at the vertices, which is given by 
\begin{equation}\label{eq:upsilon_ij}
    \begin{split}
        \Upsilon^{ij}(\iu\Omega_m,\v{p}) &=  T \sum_{n \in \ZZ} \int \frac{\D^3 k}{(2\pi)^3} \frac{k^i k^j}{m^2} G_{0}(k+p/2) G_{0}(k-p/2) \\
        &= \int \frac{\D^3 k}{(2\pi)^3} \frac{k^i k^j}{m^2} \frac{n(\v{k}-\v{p}/2) - n(\v{k}+\v{p}/2)}{\xi(\v{k}-\v{p}/2) - \xi(\v{k}+\v{p}/2) + \iu \Omega_m}.
    \end{split}
\end{equation}
This tensor response function can be split into its transverse and longitudinal parts,
\begin{equation}
    \Upsilon^{ij}(p) = \left( \delta^{ij} - \frac{p^i p^j}{\v{p}^2} \right) \Upsilon_{\perp}(p) + \frac{p^i p^j}{\v{p}^2} \Upsilon_{\parallel}(p),
\end{equation}
which in turn can be computed separately. 
In the end, we will perform an angular average over this tensor, meaning that we are ultimately interested in the function
\begin{equation}
    \delta^{ij} \tilde{\Upsilon}(p) = \delta^{ij} \left[ \frac{2}{3} \Upsilon_{\perp}(p) + \frac{1}{3} \Upsilon_{\parallel}(p) \right]. 
\end{equation}
The functions $\Upsilon_{\perp,\parallel}(p)$ are found to be
\begin{equation}
    \begin{split}
        \Upsilon_{\perp}(p) = -\frac{\nu(0)}{8 \hat{p}} v_F^2 \Bigg\{ &\frac{1}{2} \left( \hat{p} + \frac{\hat{\omega}}{\hat{p}} \right) \left[ \frac{5}{3} - \left( \hat{p} +  \frac{\hat{\omega}}{\hat{p}} \right)^2 \right] \\
        + &\frac{1}{4} \left[ 1 - \left( \hat{p} +  \frac{\hat{\omega}}{\hat{p}} \right)^2 \right]^2 \log\left(\frac{\hat{p} + \hat{\omega}/\hat{p} + 1}{\hat{p} + \hat{\omega}/\hat{p} - 1}\right) \Bigg\} + (\hat{p} \leftrightarrow - \hat{p}, \hat{\omega} \leftrightarrow - \hat{\omega}),
    \end{split}
\end{equation}
and
\begin{equation}
    \begin{split}
        \Upsilon_{\parallel}(p) = &-\frac{\nu(0)}{4 \hat{p}} v_F^2 \Bigg\{  \frac{2}{3} \left( \hat{p} - \frac{\hat{\omega}}{\hat{p}} \right)  \\
        &+  \left( \frac{\hat{\omega}}{\hat{p}} \right)^2 \left[ \left( \hat{p} +  \frac{\hat{\omega}}{\hat{p}} \right) + \frac{1}{2}\left[ 1 - \left( \hat{p} +  \frac{\hat{\omega}}{\hat{p}} \right)^2 \right] \log\left(\frac{\hat{p} + \hat{\omega}/\hat{p} + 1}{\hat{p} + \hat{\omega}/\hat{p} - 1}\right)  \right]  \Bigg\} + (\hat{p} \leftrightarrow - \hat{p}, \hat{\omega} \leftrightarrow - \hat{\omega}).
    \end{split}
\end{equation}
The imaginary part of the function $\tilde{\Upsilon}(\omega + \iu 0, \v{p})$ is shown on the right-hand side of Fig.~\ref{fig:ph_bubbles}.

Finally, we will also encounter the particle-hole bubble with a single such velocity insertion
\begin{equation}
    \Upsilon^{i}(p) = \frac{1}{\beta V} \sum_{k} \frac{k^i}{m} G_{0}(k-p/2) G_{0}(k+p/2).
\end{equation}
By making a tensor decomposition, $\Upsilon^{i}(p) = \frac{p^i}{\abs{\v{p}}} \chi(p)$ we can instead compute the scalar bubble
\begin{equation}
    \chi(p) = \frac{1}{\abs{\v{p}}} \frac{1}{\beta V} \sum_{k} \frac{\v{p} \cdot \v{k}}{m} G_{0}(k-p/2) G_{0}(k+p/2).
\end{equation}
Now we observe that $\v{k}\cdot \v{p} /m = \xi(\v{k}+\v{p}/2) - \xi(\v{k}-\v{p}/2) = p_0 - \left[G_{0}^{-1}(k+p/2) - G_{0}^{-1}(k-p/2)\right]$.
Inserting this relation into the expression for $\chi(p)$ immediately lets us conclude that 
\begin{equation}\label{eq:Upsilon_i_simplified}
    \Upsilon^{i}(p) = \frac{p^i}{\abs{\v{p}}} \frac{p_0}{\abs{\v{p}}} \frac{1}{\beta V} \sum_{k} G_{0}(k-p/2) G_{0}(k+p/2) = \frac{p^i p_{0}}{\v{p}^2} \Pi(p).
\end{equation}

\subsubsection{Imaginary part}

The imaginary part of the particle-hole bubbles can be computed directly by performing the analytical continuation $\iu\Omega_m \to \omega + \iu 0$ and using the Dirac identity $1/(x+\iu 0) = \mathrm{P}(1/x) - \iu \pi \delta(x)$.
This yields
\begin{equation}
    \Im\left[ \Pi(\omega+\iu0,\v{p})\right] = - \pi \int \frac{\D^3 k}{(2\pi)^3} \left[ n(\v{k}-\v{p}) - n(\v{k}) \right]\delta\left( \xi(\v{k}-\v{p}) - \xi(\v{k}) + \omega \right).
\end{equation}
Computing the resulting integral in a Fermi-surface average and assuming that $p/2k_F \ll 1$ and $\omega/(v_F p) < 1$ we find the well-known result \cite{Hertz-Hertz-QuantumCriticalPhenomena-1976}
\begin{equation}\label{eq:im_Pi}
    \Im\left[ \Pi(\omega+\iu0,\v{p})\right] \simeq - \frac{\pi}{2} \nu(0) \frac{\omega}{v_F p}.
\end{equation}
Repeating the same calculation for the function $\tilde{\Upsilon}(p) = \Upsilon_{\parallel}(p)/3 + 2 \Upsilon_{\perp}(p)/3$ we find that 
\begin{equation}\label{eq:im_Upsilon_tilde}
    \Im\left[ \tilde{\Upsilon}(\omega+\iu0,\v{p})\right] \simeq - \frac{v_F^2}{3} \frac{\pi}{2} \nu(0) \frac{\omega}{v_F p}.
\end{equation}
At this point we note that the leading terms of Eq.~\eqref{eq:Upsilon_i_simplified} and Eq.~\eqref{eq:im_Upsilon_tilde} in an expansion in $\omega/(v_F p)$ have the same functional form as the imaginary part of the normal particle-hole bubble in Eq.~\eqref{eq:im_Pi}. 
This will become important momentarily. 

\subsubsection{Temperature dependence}

By letting $\hat{\omega} \to 0 $ and $\hat{p}\to 0$ in Eq.~\eqref{eq:Pi(p)} we see that $\Pi(0) = - \nu(0)$ and consequently that the zero-temperature squared mass of the propagator $D(q)$ is given by $ g_c - g \equiv 1 - J \nu(0)/2$.
For $g = g_c$, the mass is instead proportional to the temperature.
This can be derived as follows.
The particle-hole bubble at finite temperature and $p=0$ is given by
\begin{align}
    \Pi(0;T) &= \int \frac{\D^3 k}{(2\pi)^3} \frac{\partial}{\partial \xi} \frac{1}{\e^{\beta\xi} + 1} \biggr\lvert_{\xi = \xi(\v{k})} = \nu(0) \frac{1}{\sqrt{\beta \epsilon_{F}}} \int_{- \mu}^{\infty} \D \xi \sqrt{\beta \left( \xi + \mu \right)} \frac{\partial }{\partial \xi} \frac{1}{\e^{\beta\xi} + 1},
\end{align}
where we have used that $\nu(\xi) = \nu(0) \sqrt{(\xi + \mu)/\epsilon_{F}}$ in $d=3$.
Using the integral representation of the polylogarithm \cite{Wood-Wood-ComputationPolylogarithms-1992}
\begin{equation}\label{eq:polylog}
    - \mathrm{Li}_{s}(-z) = \frac{1}{\Gamma(s)} \int_{0}^{\infty} \D t \frac{t^{s-1}}{\e^{t}/z + 1},
\end{equation}
we find that
\begin{equation}
    \Pi(0;T) = \frac{\nu(0)}{2} \frac{\Gamma(1/2)}{\sqrt{\beta \epsilon_{F}}} \mathrm{Li}_{1/2} \left( - \e^{\beta \mu} \right).
\end{equation}
The dependence on $\epsilon_F$ can be eliminated by taking into account the drift of the chemical potential at finite temperature $T$, which is done by equating the mean-field zero-temperature particle density $2 \nu(0) \epsilon_{F}/3$ with the (mean-field) particle density at finite temperature \cite{Mayrhofer-Chubukov-StonerTransitionFinite-2025}.
In total this yields
\begin{equation}\label{eq:Pi(T)}
    \Pi(0;T) = \nu(0) \frac{\Gamma(1/2)}{2} \mathrm{Li}_{1/2}\left( - \e^{\beta \mu} \right)  \left[ - \frac{3}{2} \Gamma(3/2) \mathrm{Li}_{3/2} \left( - \e^{\beta \mu} \right) \right]^{-1/3} .
\end{equation}
This function admits an expansion in $T/\mu$ which can be obtained by applying the Sommerfeld expansion \cite{Landau-Lifshitz-StatisticalPhysicsPart-1976} to Eq.~\eqref{eq:polylog}. 
This yields
\begin{equation}
    \mathrm{Li}_{s}(-\e^{x}) \overset{x\to\infty}{\sim} - \frac{x^s}{\Gamma(s+1)} \left[ 1 + \frac{\pi^2}{6} s (s-1) \frac{1}{x^2} \right].
\end{equation}
Applied to Eq.~\eqref{eq:Pi(T)}, we find
\begin{equation}\label{eq:Pi_T_expansion}
    \Pi(0;T) \approx - \nu(0) \left[ 1 - \frac{\pi^2}{12} \left( \frac{T}{\epsilon_F} \right)^2 \right].
\end{equation}
Hence, the QCP can be approached by tuning $g \to g_{c}$ at zero temperature, or lowering the temperature $T\to 0$ at $g = g_{c}$.

\section{Fluctuation response kernel}\label{suppl:response}

The fluctuation response kernel is found by expanding the fluctuation action $S_{\mathrm{fluc}}[A] = \tr \log (-\beta D^{-1}[A])$ to second order in  the vector potential $A_{\mu}$.
The $A_{\mu}$-dependence of $D^{-1}$ is inherited from that of $\mathcal{G}^{-1}$, which in turn is obtained by letting $\partial_{\tau} \mapsto \partial_{\tau} - \iu e_{*} A_{0} \sigma^{3}/2$ and $- \iu \bm{\nabla} \mapsto - \iu \bm{\nabla} + e_{*} \v{A} \sigma^{3}/2$.
Here, $e_{*} \equiv e/(m c)$ denotes the effective charge of the fermions under the spin gauge field.  
As derived in the manuscript, the fluctuation response kernel is given by
\begin{equation}\label{eq:Q_fluc_supp}
        Q^{\mu\nu}_{\mathrm{fluc}}(x,x') = \frac{\delta^2 \tr \log \left(- \beta D^{-1}[A]\right)}{\delta A_{\mu}(x) \delta A_{\nu}(x')} \biggr\lvert_{A = 0}.
    \end{equation}
    This yields 
    \begin{subequations}
        \begin{align}
                Q^{\mu\nu}_{\mathrm{fluc}}(x,x') = &- \int \prod_{i=1}^{4} \D z_{i} \,  D(z_1,z_2) \Lambda^{\mu}(z_2,z_3;x) D (z_3,z_4) \Lambda^{\nu}(z_4,z_1;x')  \label{eq:AL_realspace_supp} \\
        &+ \int \prod_{i=1}^{2} \D z_{i} \, D(z_1,z_2) \Xi^{\mu\nu}(z_2,z_1;x,x'), \label{eq:MT_DOS_DIA_realspace_supp}
    \end{align}
    \end{subequations}
    where the effective vertices are 
    \begin{subequations}
        \begin{align}
            \Lambda^{\mu}(z_1,z_2;x) &\coloneqq \frac{\delta D^{-1}[A](z_1,z_2)}{\delta A_{\mu}(x)} \biggr\lvert_{A = 0} \label{eq:three_pt_realspace_supp}\\
            \Xi^{\mu\nu}(z_1,z_2;x,x') &\coloneqq \frac{\delta^2 D^{-1}[A](z_1,z_2)}{\delta A_{\mu}(x) \delta A_{\nu}(x')} \biggr\lvert_{A = 0}. \label{eq:four_pt_realspace_supp}
        \end{align}
    \end{subequations}

\begin{figure}[htb]
    \centering
    \includegraphics[width=0.6\linewidth]{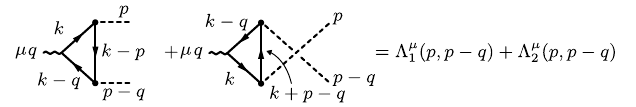}
    \caption{The two different triangle vertices contributing to the AL diagrams.}
    \label{fig:triangles}
\end{figure}

The expressions for $\Lambda^{\mu}(z_1,z_2;x)$ and $\Xi^{\mu\nu}(z_1,z_2;x,x')$ are now derived as follows.
Performing one functional derivative of the propagator in Eq.~\eqref{eq:D_traceformula} yields
\begin{align}
    \frac{\delta D^{-1}[A](z_1,z_2)}{\delta A_{\mu}(x)} &= - \frac{1}{4} \frac{\delta }{\delta A_{\mu}(x)} \tr \left( \mathcal{G}_{0}[A](z_2,z_1) \sigma^{+} \mathcal{G}_{0}[A](z_1,z_2) \sigma^{-} \right) \notag \\
    \begin{split}
         &= \frac{1}{4} \int \prod_{j=1}^{2} \D y_j  \tr \left( \mathcal{G}_{0}[A](z_2,y_1) \frac{\delta \mathcal{G}^{-1}_{0}[A](y_1,y_2)}{\delta A_{\mu}(x)} \mathcal{G}_{0}[A](y_2,z_1) \sigma^{+} \mathcal{G}_{0}[A](z_1,z_2) \sigma^{-} \right) \\
         &\,+ \frac{1}{4} \int \prod_{j=1}^{2} \D y_j \tr \left( \mathcal{G}_{0}[A](z_2,z_1) \sigma^{+} \mathcal{G}_{0}[A](z_1,y_1) \frac{\delta \mathcal{G}^{-1}_{0}[A](y_1,y_2)}{\delta A_{\mu}(x)}  \mathcal{G}_{0}[A](y_2,z_2) \sigma^{-} \right). \label{eq:D_one_derivative}
    \end{split}
\end{align}
To obtain the three-point vertex, we now let $A \to 0$ and use that
\begin{equation}\label{eq:current_vertex}
    \frac{\delta \mathcal{G}^{-1}_{0}[A](y_1,y_2)}{\delta A_{\mu}(x)}\biggr\lvert_{A = 0} = e_{*}\frac{\sigma^{3}}{2} \gamma^{\mu}(y_1,y_2;x),
\end{equation}
where $\gamma^{\mu}(y_1,y_2;x)$ denotes the normal current vertex \cite{Schrieffer-Schrieffer-TheorySuperconductivity-1999}.
This yields
\begin{equation}\label{eq:AL_vertex_realspace}
    \Lambda^{\mu}(z_1,z_2;x) = \frac{e_{*}}{8} \int \prod_{j=1}^{2} \D y_j \gamma^{\mu}(y_1,y_2;x) \Big[ G_{0}(z_2,y_1) G_{0}(y_2,z_1) G_{0}(z_1,z_2) - G_{0}(z_2,z_1) G_{0}(z_1,y_1) G_{0}(y_2,z_2) \Big].
\end{equation}
Using the Fourier transform of the current vertex \cite{Schrieffer-Schrieffer-TheorySuperconductivity-1999}
\begin{equation}
    \gamma^{\mu}(y_1,y_2;x) = \frac{1}{(\beta V)^2} \sum_{k q} \e^{\iu k \cdot (y_1 - y_2) + \iu q \cdot (y_2 - x)} \gamma^{\mu}(k,k-q),
\end{equation}
we find 
\begin{equation}
    \Lambda^{\mu}(z_1,z_2;x) = \frac{1}{(\beta V)^2} \sum_{p q} \e^{\iu p \cdot (z_1 - z_2) + \iu q \cdot (z_2 - x)} \Lambda^{\mu}(p,p-q),
\end{equation}
where the momentum-space vertex has been introduced as 
\begin{equation}
    \Lambda^{\mu}(p,p-q) = \frac{e_{*}}{8} \frac{1}{\beta V} \sum_{k} G_{0}(k) \gamma^{\mu}(k,k-q) G_{0}(k-q) \Big[ G_{0}(k+p-q) - G_{0}(k-p) \Big].
\end{equation}
The two contributions to the three-point vertex are illustrated in Fig.~\ref{fig:triangles}.

\begin{figure}[htb]
    \centering
    \includegraphics[width=0.7\linewidth]{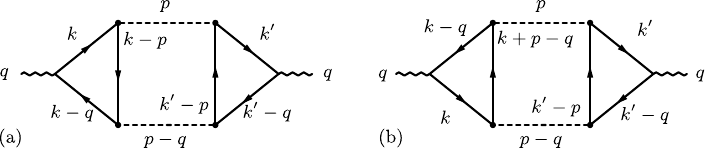}
    \caption{ The two inequivalent AL diagrams with momentum labels.}
    \label{fig:AL_diagrams}
\end{figure}

The two diagrams that contribute to $\Lambda^{\mu}(z_1,z_2;x)$ give rise to four AL diagrams that are pairwise topologically equivalent. 
Using Eq.~\eqref{eq:AL_realspace_supp} and Eq.~\eqref{eq:AL_vertex_realspace}, we can transform the resulting expression for the AL diagrams to momentum space and obtain
\begin{align}
    Q_{\mathrm{AL}}^{\mu\nu}(x,x') &= - \int \prod_{i=1}^{4} \D z_{i} \, D(z_1,z_2) \Lambda^{\mu}(z_2,z_3;x) D(z_3,z_4) \Lambda^{\nu}(z_4,z_1;x') \notag \\
    &= \frac{1}{\beta V} \sum_{q} \e^{\iu q \cdot(x-x')} Q_{\mathrm{AL}}^{\mu\nu}(q),
\end{align}
where
\begin{align}
    Q^{\mu\nu}_{\mathrm{AL}}(q) &= - \frac{1}{\beta V} \sum_{p} \Lambda^{\mu}(p-q,p) D(p) \Lambda^{\nu}(p,p-q) D(p-q) \notag \\
    \begin{split}
        &= - \frac{e_{*}^2 }{64} \frac{1}{(\beta V)^3} \sum_{p k k'} D(p) D(p-q) G_{0}(k) \gamma^{\mu}(k,k-q) G_{0}(k-q) \left[ G_{0}(k+p-q) - G_{0}(k-p) \right] \\
        & \hspace{13em}\times G_{0}(k') \gamma^{\nu}(k',k'-q) G_{0}(k'-q) \left[ G_{0}(k'+p-q) - G_{0}(k'-p) \right].
    \end{split}
\end{align}
The momentum-space representations of the AL diagrams are illustrated graphically in Fig.~\ref{fig:AL_diagrams}.

To obtain the four-point vertex, we perform another functional derivative of Eq.~\eqref{eq:D_one_derivative}.
This yields
\begin{subequations}
    \begin{align}
        \frac{\delta^2 D^{-1}[A](z_1,z_2)}{\delta A_{\mu}(x) \delta A_{\nu}(x')}&  \notag \\
        = - \frac{1}{4} \int \prod_{\ell = 1}^{4} \D y_{\ell} \tr \Bigg[ &\mathcal{G}_{0}[A](z_2,y_3) \frac{\delta \mathcal{G}^{-1}_{0}[A](y_3,y_4)}{\delta A_{\mu}(x)} \mathcal{G}_{0}[A](y_4,y_1) \frac{\delta \mathcal{G}^{-1}_{0}[A](y_1,y_2)}{\delta A_{\nu}(x')} \mathcal{G}_{0}[A](y_2,z_1) \sigma^{+} \mathcal{G}_{0}[A](z_1,z_2) \sigma^{-} \\
        +\,&\mathcal{G}_{0}[A](z_2,y_1) \frac{\delta \mathcal{G}^{-1}_{0}[A](y_1,y_2)}{\delta A_{\nu}(x')} \mathcal{G}_{0}[A](y_2,y_3) \frac{\delta \mathcal{G}^{-1}_{0}[A](y_3,y_4)}{\delta A_{\mu}(x)} \mathcal{G}_{0}[A](y_4,z_1) \sigma^{+} \mathcal{G}_{0}[A](z_1,z_2) \sigma^{-}  \\
        +\,&\mathcal{G}_{0}[A](z_2,y_1) \frac{\delta \mathcal{G}^{-1}_{0}[A](y_1,y_2)}{\delta A_{\nu}(x')} \mathcal{G}_{0}[A](y_2,z_1) \sigma^{+} \mathcal{G}_{0}[A](z_1,y_3) \frac{\delta \mathcal{G}^{-1}_{0}[A](y_3,y_4)}{\delta A_{\mu}(x)} \mathcal{G}_{0}[A](y_4,z_2) \sigma^{-} \\
        +\,&\mathcal{G}_{0}[A](z_2,y_3) \frac{\delta \mathcal{G}^{-1}_{0}[A](y_3,y_4)}{\delta A_{\mu}(x)} \mathcal{G}_{0}[A](y_4,z_1) \sigma^{+} \mathcal{G}_{0}[A](z_1,y_1) \frac{\delta \mathcal{G}^{-1}_{0}[A](y_1,y_2)}{\delta A_{\nu}(x')} \mathcal{G}_{0}[A](y_2,z_2) \sigma^{-} \\
        +\,&\mathcal{G}_{0}[A](z_2,z_1) \sigma^{+} \mathcal{G}_{0}[A](z_1,y_3) \frac{\delta \mathcal{G}^{-1}_{0}[A](y_3,y_4)}{\delta A_{\mu}(x)} \mathcal{G}_{0}[A](y_4,y_1)  \frac{\delta \mathcal{G}^{-1}_{0}[A](y_1,y_2)}{\delta A_{\nu}(x')} \mathcal{G}_{0}[A](y_2,z_2) \sigma^{-} \\
        +\,&\mathcal{G}_{0}[A](z_2,z_1) \sigma^{+} \mathcal{G}_{0}[A](z_1,y_1) \frac{\delta \mathcal{G}^{-1}_{0}[A](y_1,y_2)}{\delta A_{\nu}(x')} \mathcal{G}_{0}[A](y_2,y_3)  \frac{\delta \mathcal{G}^{-1}_{0}[A](y_3,y_4)}{\delta A_{\mu}(x)} \mathcal{G}_{0}[A](y_4,z_2) \sigma^{-} \Bigg] \\
        + \frac{1}{4} \int \prod_{\ell = 1}^{2} \D y_{\ell} \tr \Bigg[ &\mathcal{G}_{0}[A](z_2,y_1) \frac{\delta^2 \mathcal{G}^{-1}_{0}[A](y_1,y_2)}{\delta A_{\mu}(x) \delta A_{\nu}(x')} \mathcal{G}_{0}[A](y_2,z_1) \sigma^{+} \mathcal{G}_{0}[A](z_1,z_2) \sigma^{-}  \\
        +\,&\mathcal{G}_{0}[A](z_2,z_1) \sigma^{+} \mathcal{G}_{0}[A](z_1,y_1) \frac{\delta^2 \mathcal{G}^{-1}_{0}[A](y_1,y_2)}{\delta A_{\mu}(x) \delta A_{\nu}(x')}  \mathcal{G}_{0}[A](y_2,z_2) \sigma^{-} \Bigg]
\end{align}
\end{subequations}
Note that the two terms that gave rise to the two manifestly different contributions to $\Lambda^{\mu}(z_1,z_2;x)$ turn out to give equivalent contributions to the four-point function $\Xi^{\mu\nu}(z_1,z_2;x,x')$ in the end.
Performing the remaining matrix trace and using the definition of the current vertex in Eq.~\eqref{eq:current_vertex} as well as
\begin{equation}\label{eq:dia_current_vertex}
    \frac{\delta^2 \mathcal{G}^{-1}_{0}[A](y_1,y_2)}{\delta A_{\mu}(x)\delta A_{\nu}(x')}\biggr\lvert_{A = 0} = e_{*}^2 \frac{\mathbf{1}}{4} \gamma^{\mu\nu}(y_1,y_2;x,x'),
\end{equation}
we find that
\begin{align}
    \Xi^{\mu\nu}(z_1,z_2;x,x') = \frac{e_{*}^2}{8} \int \prod_{\ell=1}^{4} \D y_{\ell} \Big[ &G_0(z_2,y_1) \gamma^{\mu}(y_1,y_2;x) G_{0}(y_2,z_1) G_{0}(z_1,y_3) \gamma^{\nu}(y_3,y_4;x') G_{0}(y_4,z_2) \label{eq:XI_1} \\
   \phantom{\prod_{\ell}^{4}}  - & G_0(z_2,y_1) \gamma^{\mu}(y_1,y_2;x) G_0(y_2,y_3) \gamma^{\nu}(y_3,y_4;x') G_0(y_4,z_1) G_0(z_1,z_2)  \label{eq:XI_2} \\
    \phantom{\prod_{\ell}^{4}} - & G_0(z_2,y_1) \gamma^{\nu}(y_1,y_2;x') G_0(y_2,y_3) \gamma^{\mu}(y_3,y_4;x) G_0(y_4,z_1) G_0(z_1,z_2)
    \Big] \label{eq:XI_3} \\
    + \frac{e_{*}^2}{8} \int \prod_{\ell=1}^{2} \D y_{\ell}  \Big[ &  G_{0}(z_1,y_1) \gamma^{\mu\nu}(y_1,y_2;x,x') G_{0}(y_2,z_2) G_{0}(z_2,z_1) \Big] .  \label{eq:XI_4}
\end{align}
After convoluting with the spin-fluctuation propagator, Eq.~\eqref{eq:XI_1} yields the Maki-Thompson (MT) diagram, Eq.~\eqref{eq:XI_2} and \eqref{eq:XI_3} yield the self-energy (or DOS) diagrams and Eq.~\eqref{eq:XI_4} the diamagnetic (DIA) diagram.
In what follows, we will convert these diagrams to their momentum-space representations. 

\begin{figure}[htb]
    \centering
    \includegraphics[width=0.4\linewidth]{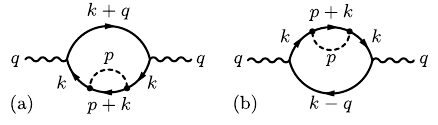}
    \caption{The two different DOS diagrams with momentum labels.}
    \label{fig:DOS_diagrams}
\end{figure}

The DOS diagram is obtained by using Eq.~\eqref{eq:XI_2} and \eqref{eq:XI_3} together with Eq.~\eqref{eq:MT_DOS_DIA_realspace_supp} and transforming to momentum space
\begin{align}
    Q^{\mu\nu}_{\mathrm{DOS}}(x,x') &= \int \prod_{i=1}^{2} \D z_i \, D(z_2,z_1) \Xi^{\mu\nu}_{\mathrm{DOS}}(z_1,z_2;x,x') \notag \\
    \begin{split}
        &= -\frac{e_{*}^2}{8} \int \prod_{i=1}^{2} \D z_i \int \prod_{j=1}^{4} \D y_{j} \, D(z_2,z_1) \Big[ G_0(z_2,y_1) \gamma^{\mu}(y_1,y_2;x) G_0(y_2,y_3) \gamma^{\nu}(y_3,y_4;x') G_0(y_4,z_1) G_0(z_1,z_2) \\
     & \hspace{15.2em} +  G_0(z_2,y_1) \gamma^{\nu}(y_1,y_2;x') G_0(y_2,y_3) \gamma^{\mu}(y_3,y_4;x) G_0(y_4,z_1) G_0(z_1,z_2) \Big] \notag
    \end{split} \\
    &= \frac{1}{\beta V} \sum_{q} \e^{\iu q \cdot (x-x')} Q^{\mu\nu}_{\mathrm{DOS}}(q),
\end{align}
where
\begin{equation}
    \begin{split}
        Q^{\mu\nu}_{\mathrm{DOS}}(q) &= - \frac{e_{*}^2}{8} \frac{1}{(\beta V)^2} \sum_{kp} D(p) \left[G_{0}(k)\right]^2 G_{0}(k+p) \Big[ \gamma^{\mu}(k,k+q) G_{0}(k+q) \gamma^{\nu}(k+q,k) \\
        &\hspace{17.6em} +  \gamma^{\mu}(k,k-q) G_{0}(k-q) \gamma^{\nu}(k-q,k) \Big]
    \end{split}
\end{equation}
The momentum-space representations of the DOS diagrams are illustrated graphically in Fig.~\ref{fig:DOS_diagrams}.

\begin{figure}[htb]
    \centering
    \begin{minipage}{0.49\textwidth}
        \includegraphics[width=0.4\linewidth]{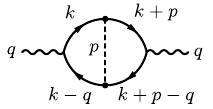}
        \caption{The MT diagram with momentum labels.}
        \label{fig:MT_diagram}
    \end{minipage}
    \hfill
    \begin{minipage}{0.49\textwidth}
        \includegraphics[width=0.22\linewidth]{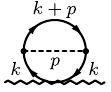}
    \caption{The DIA diagram with momentum labels.}
    \label{fig:DIA_diagram}
    \end{minipage}
\end{figure}

Similarly, the MT diagram is now obtained by using Eq.~\eqref{eq:XI_1} together with Eq.~\eqref{eq:MT_DOS_DIA_realspace_supp} and transforming to momentum space
\begin{align}
    Q^{\mu\nu}_{\mathrm{MT}}(x,x') &= \int \prod_{i=1}^{2} \D z_i \, D(z_2,z_1) \Xi^{\mu\nu}_{\mathrm{MT}}(z_1,z_2;x,x') \notag \\
    &= \frac{e_{*}^2}{8} \int \prod_{i=1}^{2} \D z_i  \int \prod_{j=1}^{4} \D y_{j} \,  D(z_2,z_1) G_0(z_2,y_1) \gamma^{\mu}(y_1,y_2;x) G_{0}(y_2,z_1) G_{0}(z_1,y_3) \gamma^{\nu}(y_3,y_4;x') G_{0}(y_4,z_2) \notag \\
    &= \frac{1}{\beta V} \sum_{q} \e^{\iu q \cdot (x-x')} Q^{\mu\nu}_{\mathrm{MT}}(q),
\end{align}
where
\begin{equation}
    Q^{\mu\nu}_{\mathrm{MT}}(q) = \frac{e_{*}^2}{8} \frac{1}{(\beta V)^2} \sum_{kp} D(p) G_{0}(k-q) \gamma^{\mu}(k-q,k) G_{0}(k) G_{0}(k+p) \gamma^{\nu}(k+p,k+p-q) G_{0}(k+p-q).
\end{equation}
The momentum-space representation of the MT diagram is illustrated graphically in Fig.~\ref{fig:MT_diagram}.

Finally, we obtain the DIA diagram by using Eq.~\eqref{eq:XI_4} together with Eq.~\eqref{eq:MT_DOS_DIA_realspace_supp} and
\begin{equation}
    \gamma^{\mu\nu}(y_1,y_2;x,x') = - \delta(y_1-y_2) \delta(y_1-x) \delta(x-x') \frac{1}{m} \delta^{\mu\nu} \left(1 - \delta^{\nu 0} \right).
\end{equation}
This yields the (momentum-independent) kernel
\begin{align}
    Q^{\mu\nu}_{\mathrm{DIA}}(x,x') &= \int \prod_{i=1}^{2} \D z_i \, D(z_2,z_1) \Xi^{\mu\nu}_{\mathrm{DIA}}(z_1,z_2;x,x') \notag \\
    &= \frac{e_{*}^2}{8} \int \prod_{i=1}^{2} \D z_i  \int \prod_{j=1}^{2} \D y_{j} \,  D(z_2,z_1) G_{0}(z_1,y_1) \gamma^{\mu\nu}(y_1,y_2;x,x') G_{0}(y_2,z_2) G_{0}(z_2,z_1) \notag \\
    &= \delta(x - x') Q^{\mu\nu}_{\mathrm{DIA}}(q),
\end{align}
where
\begin{equation}
    Q^{\mu\nu}_{\mathrm{DIA}}(q) = - \frac{e_{*}^2}{8 m} \delta^{\mu\nu} \left( 1 - \delta^{\nu 0} \right) \frac{1}{(\beta V)^2} \sum_{kp} D(p) \left[ G_{0}(k) \right]^2 G_{0}(k+p).
\end{equation}
The momentum-space representation of the DIA diagram is illustrated graphically in Fig.~\ref{fig:DIA_diagram}.

\subsection{Proof of gauge invariance}\label{suppl:gauge_inv}

The linear-response kernel $Q^{\mu\nu}(q)$ is constrained by the conservation law for the $z$ component of the spin density.
In the gauge-theory formulation of the linear-response theory, this is equivalent to gauge-invariance of the effective action $S_{\mathrm{eff}}[A]$ under $A_{\mu} \mapsto A_{\mu} - \partial_{\mu} \chi$ \cite{Altland-Simons-CondensedMatterField-2023}.
In momentum space, the corresponding Ward identity reads $q_{\mu} Q^{\mu\nu}(q) = 0$.
We will now demonstrate this identity for the set of fluctuation diagrams that derive from the Gaussian-fluctuation approximation. 
This calculation parallels that of Ref.~\cite{Boyack-Boyack-RestoringGaugeInvariance-2018} for the pairing channel, but here it applies to the spin channel.

For this proof, we will make use of the bare Ward-Takahashi identity \cite{Schrieffer-Schrieffer-TheorySuperconductivity-1999}
\begin{equation}\label{eq:bare_Ward}
    q_{\mu} \gamma^{\mu}(k-q,k) = G_{0}^{-1}(k) - G_{0}^{-1}(k-q).
\end{equation}
Considering the contraction of the AL diagrams first, we obtain
\begin{align}
    \begin{split}
        q_{\mu}Q^{\mu\nu}_{\mathrm{AL}}(q)
        &= - \frac{e_{*}^2 }{64} \frac{1}{(\beta V)^3} \sum_{p k k'} D(p) D(p-q) G_{0}(k) q_{\mu}\gamma^{\mu}(k,k-q) G_{0}(k-q) \left[ G_{0}(k+p-q) - G_{0}(k-p) \right] \\
        & \hspace{13em}\times G_{0}(k') \gamma^{\nu}(k',k'-q) G_{0}(k'-q) \left[ G_{0}(k'+p-q) - G_{0}(k'-p) \right] \notag
    \end{split} \\
    \begin{split}
        &= - \frac{e_{*}^2 }{64} \frac{1}{(\beta V)^3} \sum_{p k k'} D(p) D(p-q) \left[ G_{0}(k-q) - G_{0}(k) \right] \left[ G_{0}(k+p-q) - G_{0}(k-p) \right] \\
        & \hspace{13em}\times G_{0}(k') \gamma^{\nu}(k',k'-q) G_{0}(k'-q) \left[ G_{0}(k'+p-q) - G_{0}(k'-p) \right].
    \end{split} 
\end{align}
By performing the summation over $k$ we can recover differences between inverses of the spin-fluctuation propagators as defined in Eq.~\eqref{eq:Dinv}.
This yields
\begin{align}
    \begin{split}
        q_{\mu}Q^{\mu\nu}_{\mathrm{AL}}(q)
        &= - \frac{e_{*}^2 }{8} \frac{1}{(\beta V)^2} \sum_{p k'} D(p) D(p-q) \left[ D^{-1}(p-q) - D^{-1}(p) \right]  \\
        & \hspace{13em}\times G_{0}(k') \gamma^{\nu}(k',k'-q) G_{0}(k'-q) \left[ G_{0}(k'+p-q) - G_{0}(k'-p) \right] \notag 
    \end{split} \\
    \begin{split}
        &= - \frac{e_{*}^2 }{8} \frac{1}{(\beta V)^2} \sum_{p k} D(p) G_{0}(k) \gamma^{\nu}(k,k-q) G_{0}(k-q)  \\
        & \hspace{13em}\times \left[ G_{0}(k-p-q) - G_{0}(k+p) - G_{0}(k-p) + G_{0}(k+p-q) \right].
    \end{split} 
\end{align}
The contraction of the MT diagram is
\begin{align}
    q_{\mu} Q^{\mu\nu}_{\mathrm{MT}}(q) &= \frac{e_{*}^2}{8} \frac{1}{(\beta V)^2} \sum_{kp} D(p) G_{0}(k-q) q_{\mu}\gamma^{\mu}(k-q,k) G_{0}(k) G_{0}(k+p) \gamma^{\nu}(k+p,k+p-q) G_{0}(k+p-q) \notag \\
    &= \frac{e_{*}^2}{8} \frac{1}{(\beta V)^2} \sum_{kp} D(p) \left[ G_{0}(k-q) - G_{0}(k) \right] G_{0}(k+p) \gamma^{\nu}(k+p,k+p-q) G_{0}(k+p-q) \notag \\
    &= \frac{e_{*}^2}{8} \frac{1}{(\beta V)^2} \sum_{kp} D(p) G_{0}(k) \gamma^{\nu}(k,k-q) G_{0}(k-q) \left[ G_{0}(k-p-q) - G_{0}(k-p) \right].
\end{align}
At this stage, we see that
\begin{align}\label{eq:sum_contract_AL+MT}
    q_{\mu} Q^{\mu\nu}_{\mathrm{AL}}(q) + q_{\mu} Q^{\mu\nu}_{\mathrm{MT}}(q) = -\frac{e_{*}^2}{8} \frac{1}{(\beta V)^2} \sum_{p k} D(p) G_{0}(k) \gamma^{\nu}(k,k-q) G_{0}(k-q) \left[ G_{0}(k+p-q) - G_{0}(k+p) \right].
\end{align}
The contraction with the DOS diagrams reads
\begin{align}
    \begin{split}
        q_{\mu}Q^{\mu\nu}_{\mathrm{DOS}}(q) &= - \frac{e_{*}^2}{8} \frac{1}{(\beta V)^2} \sum_{kp} D(p) \left[G_{0}(k)\right]^2 G_{0}(k+p) \Big[ q_{\mu}\gamma^{\mu}(k,k+q) G_{0}(k+q) \gamma^{\nu}(k+q,k) \\
        &\hspace{17.6em} +  q_{\mu}\gamma^{\mu}(k,k-q) G_{0}(k-q) \gamma^{\nu}(k-q,k) \Big]
    \end{split} \notag \\
    &= - \frac{e_{*}^2}{8} \frac{1}{(\beta V)^2} \sum_{kp} D(p) \gamma^{\nu}(k,k-q) \Big[ G_{0}(k-q) - G_{0}(k) \Big] \Big[ G_{0}(k-q) G_{0}(k-q+p) + G_{0}(k) G_{0}(k+p) \Big] \notag \\
    &= - q_{\mu} Q^{\mu\nu}_{\mathrm{AL}}(q) - q_{\mu} Q^{\mu\nu}_{\mathrm{MT}}(q) - \frac{e_{*}^2}{8} \frac{1}{(\beta V)^2} \sum_{kp} D(p)  \left[ G_{0}(k) \right]^2 G_{0}(k+p) \Big[ \gamma^{\nu}(k+q,k) - \gamma^{\nu}(k,k-q) \Big],
\end{align}
where we have identified the contribution from Eq.~\eqref{eq:sum_contract_AL+MT}.
Finally, using that 
\begin{equation}
    \gamma^{\nu}(k+q,k) - \gamma^{\nu}(k,k-q) = \frac{q^{\nu}}{m} \left( 1 - \delta^{\nu 0} \right), 
\end{equation}
shows that the inclusion of the DIA diagram renders the response kernel gauge-invariant
\begin{equation}
    q_{\mu}Q^{\mu\nu}_{\mathrm{AL}}(q) + q_{\mu}Q^{\mu\nu}_{\mathrm{MT}}(q) + q_{\mu}Q^{\mu\nu}_{\mathrm{DOS}}(q) + q_{\mu}Q^{\mu\nu}_{\mathrm{DIA}}(q) = 0 .
\end{equation}

\subsection{Superfluid stiffness in the normal state}\label{suppl:rho_s}

The superfluid (or spin) stiffness can be probed by letting $\Omega_m\to0$ before taking the uniform limit $\v{q}\to 0$ in $Q^{ij}_{\mathrm{fluc}}(q)$.
To study this limit, we use the static and zero-momentum limit of Eq.~\eqref{eq:bare_Ward}
\begin{equation}\label{eq:current_vertex_id}
    G_{0}(k) \gamma^{i}(k,k) G_{0}(k) = \frac{\partial G_{0}(k)}{\partial k^{i}}.
\end{equation}
Similar to the fluctuation response in the pairing channel \cite{Boyack-Boyack-RestoringGaugeInvariance-2018} this yields
\begin{align}
    \lim_{\v{q} \to 0}\lim_{\Omega_m\to 0} Q^{ij}_{\mathrm{AL}}(q) &= -\frac{e_{*}^2}{\beta V} \sum_{p} D(p) \frac{\partial^2 D^{-1}(p)}{\partial p^{i} \partial p^{j}} \\
    \lim_{\v{q} \to 0}\lim_{\Omega_m\to 0}Q^{ij}_{\mathrm{MT}}(q) &= +\frac{e_{*}^2}{2}\frac{1}{\beta V} \sum_{p} D(p) \frac{\partial^2 D^{-1}(p)}{\partial p^{i} \partial p^{j}} \\
    \lim_{\v{q} \to 0}\lim_{\Omega_m\to 0}Q^{ij}_{\mathrm{DOS}}(q) &= -\lim_{\v{q} \to 0}\lim_{\Omega_m\to 0}Q^{ij}_{\mathrm{DIA}}(q) + \frac{e_{*}^2}{2}\frac{1}{\beta V} \sum_{p} D(p) \frac{\partial^2 D^{-1}(p)}{\partial p^{i} \partial p^{j}}.
\end{align}
This establishes that the fluctuation diagrams do not give rise to an anomalous superfluid stiffness in the normal state.

\subsection{Spin conductivity tensor}\label{suppl:sigma}

\begin{figure}[htb]
    \centering
    \includegraphics[width=0.9\linewidth]{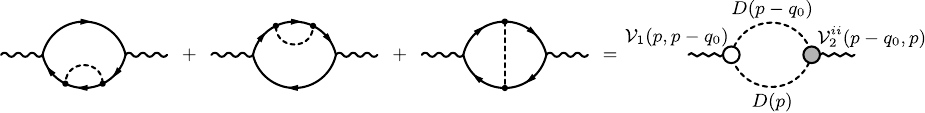}
    \caption{Schematic illustration of the rewriting of the sum of the MT and DOS diagrams as a spin-fluctuation bubble.}
    \label{fig:MTDOS_sum}
\end{figure}

We now turn to the calculation of the fluctuation spin conductivity tensor.
This is done by taking the uniform limit of the response kernel $Q^{ij}_{\mathrm{fluc}}(q)$.
Here, we will make repeated use of the identity that
\begin{equation}\label{eq:partial_fraction_identity}
    G_{0}(k) G_{0}(k-q_{0}) = \frac{1}{q_{0}} \left[ G_{0}(k-q_{0}) - G_{0}(k) \right],
\end{equation}
where $k-q_{0}$ should be understood as the four-vector $(k_{0}-q_{0},\v{k})$ \cite{Lundemo-Sudbo-FluctuationConductivityUltraclean-2026b}.
The strategy here will be to express the sum of the MT and DOS diagrams in a form comparable to the AL diagrams \cite{Lundemo-Sudbo-FluctuationConductivityUltraclean-2026b}.
Considering first the MT diagram, we find
\begin{align}
    Q^{ij}_{\mathrm{MT}}(q_{0},\v{0}) &= \frac{e_{*}^2}{8} \frac{1}{(\beta V)^2} \sum_{k p} D(p) \gamma^{i}(k,k) \gamma^{j}(k+p,k+p) \frac{1}{q_{0}} \Big[ G_{0}(k-q_{0}) - G_{0}(k) \Big] \frac{1}{q_{0}} \Big[ G_{0}(k + p -q_{0}) - G_{0}(k+p) \Big] \notag \\
    \begin{split}
        &= \frac{e_{*}^2}{8 q_{0}^2} \frac{1}{(\beta V)^2} \sum_{k p} D(p) \gamma^{i}(k,k) \gamma^{j}(k+p,k+p) \Big[ G_{0}(k-q_{0}) G_{0}(k+p-q_{0}) - G_{0}(k-q_{0}) G_{0}(k+p)   \\
        &\hspace{19.2em} - G_{0}(k) G_{0}(k+p-q_{0}) + G_{0}(k) G_{0}(k+p) \Big]
    \end{split} \notag \\
    &= \frac{e_{*}^2}{8 q_{0}^2} \frac{1}{(\beta V)^2} \sum_{k p}\gamma^{i}(k,k) \gamma^{j}(k+p,k+p) G_{0}(k) G_{0}(k+p) \Big[ 2D(p) - D(p-q_{0}) - D(p+q_{0}) \Big].
\end{align}
Similarly, the DOS diagrams simplify to 
\begin{align}
    Q^{ij}_{\mathrm{DOS}}(q_{0},\v{0}) &= -\frac{e_{*}^2}{8} \frac{1}{(\beta V)^2} \sum_{k p} D(p) \gamma^{i}(k,k) \gamma^{j}(k,k) G_{0}(k) G_{0}(k+p) \frac{1}{q_{0}} \Big[ G_{0}(k-q_{0}) - G_{0}(k+q_{0})  \Big] \notag \\
    &= -\frac{e_{*}^2}{8 q_{0}^2} \frac{1}{(\beta V)^2} \sum_{k p} D(p) \gamma^{i}(k,k) \gamma^{j}(k,k) \Big[ G_{0}(k-q_{0}) G_{0}(k+p) + G_{0}(k+q_{0}) G_{0}(k+p) - 2 G_{0}(k) G_{0}(k+p) \Big] \notag \\
    &= + \frac{e_{*}^2}{8 q_{0}^2} \frac{1}{(\beta V)^2} \sum_{k p} \gamma^{i}(k,k) \gamma^{j}(k,k) G_{0}(k) G_{0}(k+p) \Big[ 2D(p) - D(p-q_{0}) - D(p+q_{0}) \Big].
\end{align}
Adding these two contributions, we find 
\begin{equation}\label{eq:sum_MT_DOS}
    Q^{ij}_{\mathrm{MT}+\mathrm{DOS}}(q_{0},\v{0}) = \frac{e_{*}^2}{4 q_{0}^2} \frac{1}{\beta V} \sum_{p} \Phi^{ij}(p) \Big[ 2D(p) - D(p-q_{0}) - D(p+q_{0}) \Big],
\end{equation} 
where we have introduced
\begin{equation}
    \Phi^{ij}(p) \coloneqq \frac{1}{2}\frac{1}{\beta V} \sum_{k} \gamma^{i}(k,k) \left[ \gamma^{j}(k,k) + \gamma^{j}(k+p,k+p) \right] G_{0}(k) G_{0}(k+p).
\end{equation}
Since $D(p) = D(-p)$, only the symmetric part of $\Phi^{ij}(p)$ contributes in the end:
\begin{equation}
    \Upsilon^{ij}(p) \coloneqq \frac{\Phi^{ij}(p) + \Phi^{ij}(-p)}{2}.
\end{equation}
The sum of these two diagrams can now be expressed as 
\begin{align}
    Q^{ij}_{\mathrm{MT}+\mathrm{DOS}}(q_{0},\v{0}) &= \frac{e_{*}^2}{4 q_{0}^2} \frac{1}{\beta V} \sum_{p} \Upsilon^{ij}(p) \Big[ 2D(p) - D(p-q_{0}) - D(p+q_{0}) \Big] \notag \\
    &= \frac{e_{*}^2}{4 q_{0}^2} \frac{1}{\beta V} \sum_{p}  \Big[ \Upsilon^{ij}(p) - \Upsilon^{ij}(p-q_{0}) \Big] \Big[ D(p) - D(p-q_{0}) \Big] \notag \\
    &= \frac{e_{*}^2}{4 q_{0}^2} \frac{1}{\beta V} \sum_{p}  \Big[ \Upsilon^{ij}(p) - \Upsilon^{ij}(p-q_{0}) \Big] D(p) \Big[ D^{-1}(p-q_{0}) - D^{-1}(p) \Big] D(p-q_{0}) \notag \\
    &= - \frac{1}{4} \frac{1}{\beta V} \sum_{p} D(p) \mathcal{V}_{1}(p,p-q_{0}) D(p-q_{0}) \mathcal{V}_{2}^{ij}(p-q_{0},p), 
\end{align}
where the effective vertices are defined as
\begin{equation}\label{eq:MTDOS_vertices}
    \mathcal{V}_{1}(p,p-q_{0}) \coloneqq \frac{e_*}{q_{0}} \left[ D^{-1}(p) - D^{-1}(p-q_{0}) \right] \quad \text{and} \quad \mathcal{V}_{2}^{ij}(p-q_{0},p) \coloneqq \frac{e_*}{q_{0}} \left[ \Upsilon^{ij}(p) - \Upsilon^{ij}(p-q_{0}) \right].
\end{equation}
An illustration of this reformulation of the sum of the MT and DOS diagrams is shown in Fig.~\ref{fig:MTDOS_sum}.
Note that there is a partial cancellation of the MT and DOS diagrams here as well, but which differs from the corresponding one in the case of the superconductor \cite{Lundemo-Sudbo-FluctuationConductivityUltraclean-2026b}.
This is highlighted by the fact that the effective bosonic vertices are not the same.

By inserting the expression for the current vertices, we find that the definition of the function $\Upsilon^{ij}(p)$ coincides with the particle-hole bubble with current vertex insertions in Eq.~\eqref{eq:upsilon_ij}
\begin{align}
    \Upsilon^{ij}(p) \equiv \frac{\Phi^{ij}(p) + \Phi^{ij}(-p)}{2}
    &= \frac{1}{4}\frac{1}{\beta V} \sum_{k} \frac{k^i}{m} \frac{1}{m} \left[ 2 k^j + p^j\right] G_{0}(k) G_{0}(k+p) + \left( p \leftrightarrow -p \right) \notag \\
    &= \frac{1}{2} \frac{1}{\beta V} \sum_{k} \frac{k^i (k^j + p^j/2)}{m^2} G_{0}(k) G_{0}(k+p) + \left( p \leftrightarrow -p \right) \notag \\
    &= \frac{1}{\beta V} \sum_{k} \frac{k^i k^j}{m^2} G_{0}(k - p/2) G_{0}(k+p/2).
\end{align}

\begin{figure}[htb]
    \centering
    \includegraphics[width=0.65\linewidth]{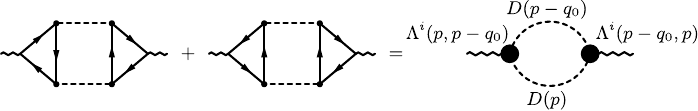}
    \caption{Schematic illustration of the AL diagrams as a spin-fluctuation bubble.}
    \label{fig:AL_sum}
\end{figure}

That the sum of the MT and DOS diagrams does not cancel the AL diagrams is apparent from bringing it to the form of a spin-fluctuation bubble diagram with different vertices.
The AL diagrams have effective vertices that can be expressed in the uniform limit as
\begin{align}
    \Lambda^{i}(p,p-q_{0}) &= \frac{e_{*}}{8} \frac{1}{\beta V } \sum_{k} \gamma^{i}(k,k) G_{0}(k) G_{0}(k-q_{0}) \Big[ G_{0}(k+p-q_{0}) - G_{0}(k-p) \Big] \notag \\
    &= \frac{e_{*}}{8 q_{0}} \frac{1}{\beta V} \sum_{k} \gamma^{i}(k,k) \Big[ G_{0}(k) G_{0}(k+p) - G_{0}(k) G_{0}(k+p-q_{0}) \Big] + \left( p \leftrightarrow -p \right).
\end{align}
These vertices therefore involve the finite-difference derivative with respect to $p_{0}$ of the particle-hole bubble with a single current-vertex insertion
\begin{equation}\label{eq:AL_upsilon_form}
    \Lambda^{i}(p,p-q_{0}) = \frac{e_{*}}{4 q_{0}} \left( \Upsilon^{i}(p) - \Upsilon^{i}(p-q_{0}) \right),
\end{equation}
where 
\begin{equation}
    \Upsilon^{i}(p) = \frac{1}{\beta V} \sum_{k} \frac{k^i}{m} G_{0}(k-p/2) G_{0}(k+p/2).
\end{equation}
The sum of the AL diagrams is illustrated schematically in Fig.~\ref{fig:AL_sum}.

% For computing the conductivity, we will use the approximation that $q_{0}\to 0$ in the vertices, while keeping the $q_{0}$-dependence of the spin-fluctuation propagators in the remaining Matsubara summation, analogous to the approximation used for the fluctuation conductivity in superconductors \cite{Larkin-Varlamov-TheoryFluctuationsSuperconductors-2005}.
% In this limit, we use the form of $\Upsilon^{i}(p)$ derived in Eq.~\eqref{eq:Upsilon_i_simplified} and get
% \begin{equation}\label{eq:Lambda_i_simplified}
%     \Lambda^{i}(p,p) = \lim_{q_{0}\to 0} \frac{e_{*}}{4 q_{0}} \left( \Upsilon^{i}(p) - \Upsilon^{i}(p-q_{0}) \right) = \frac{e_*}{4} \frac{p^i}{\v{p}^2} \left( \Pi(p) + p_{0} \frac{\partial \Pi(p)}{\partial p_0} \right). 
% \end{equation}
% Likewise, we find under the same approximation that
% \begin{equation}\label{eq:MTDOS_vertices_simplified}
%     \mathcal{V}_{1}(p,p) = e_* \frac{\partial D^{-1}(p)}{\partial p_{0}} \quad \text{and} \quad \mathcal{V}_{2}^{ij}(p,p) = e_* \frac{\partial \Upsilon^{ij}(p)}{\partial p_{0}}.
% \end{equation}

\section{Contour integral}\label{suppl:contour}

\begin{figure}[htb]
    \centering
    \includegraphics[width=0.35\linewidth]{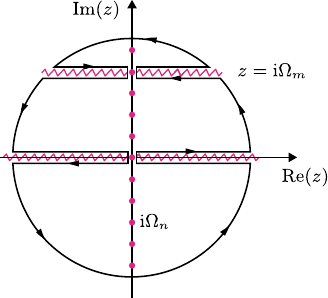}
    \caption{Integration contour that encloses the poles at the bosonic Matsubara frequencies and avoids the branch cuts at $\Im(z) = 0, \Omega_m$.
    The poles are denoted by the pink dots and the branch cuts by the pink zig-zag lines.
    }
    \label{fig:contour}
\end{figure}

The bosonic Matsubara summation that remains to obtain the spin conductivity takes the form
\begin{equation}
	J(\iu\Omega_m;\v{p}) = T \sum_{n \in \ZZ} F(\iu \Omega_n, \iu\Omega_n - \iu\Omega_m; \v{p}) = \frac{1}{4\pi \iu} \oint_{C} \D z \coth\left( \frac{z}{2T} \right) F (z,z-\iu\Omega_m; \v{p}),
\end{equation}
where the function $F(z,z';\v{p})$ is given by 
\begin{equation}
	F(z,z';\v{p}) \coloneqq D(z,\v{p}) D(z',\v{p}) V_{1}(z,z';\v{p}) V_{2}(z',z;\v{p}),
\end{equation}
and $V_{1/2}(z,z') \in \left\{ \Lambda(z,z'), \mathcal{V}_{1/2}(z,z') \right\}$ denotes the effective boson vertices.
To simplify the notation, we suppress the $\v{p}$ dependence in the following.
The contour $C$ is the contour shown in Fig.~\ref{fig:contour} including all the simple poles along the imaginary axis while avoiding the branch cuts at $\Im(z) = 0 $ and $\Im(z) = \Omega_m$ that derive from the propagators.
We find
\begin{align}
	J(\iu\Omega_m) = \frac{1}{4\pi \iu} \mathrm{P}\int_{-\infty}^{+\infty} \D z \coth\left(\frac{z}{2T}\right) \Big[ &F(z+\iu0,z-\iu\Omega_m) - F(z-\iu0, z- \iu\Omega_m)  \notag \\
	+ &F(z+\iu\Omega_m, z + \iu 0) - F(z+\iu\Omega_m, z- \iu 0) \Big],
\end{align}
where $\mathrm{P}$ denotes the principal value.
At this point, we can perform the analytical continuation $\iu\Omega_m \to \omega + \iu 0$. Introducing the notation for the analytical function above and below the cuts at $z= 0,\iu\Omega_m$ as $F^{RR}(z,z') \coloneqq F(z+\iu0,z'+\iu0)$, $F^{RA}(z,z') \coloneqq F(z+\iu0,z'-\iu0)$, etc., we obtain
\begin{equation}\label{eq:J_final}
	J(\omega + \iu 0) = \frac{1}{4\pi \iu} \mathrm{P} \int_{-\infty}^{+\infty} \D z \coth\left(\frac{z}{2T}\right) \Big[ F^{RA}(z,z-\omega) - F^{AA}(z,z-\omega) + F^{RR}(z+\omega,z) - F^{RA}(z+\omega, z) \Big].
\end{equation}

\subsection{Separation of vertex contributions}
With the result in Eq.~\eqref{eq:J_final} of the previous section at hand, we are in a position to identify the most dominant contributions to the spin conductivity in the uniform limit. 
The leading part of Eq.~\eqref{eq:AL_upsilon_form} is simply given by $\Lambda^{i}(p,p) \approx - \nu(0) e_{*}/4 \times p^i/\v{p}^2$.
From Eqs.~\eqref{eq:MTDOS_vertices} we see that the vertices $\mathcal{V}_{1/2}(p,p-q_0)$ that go into the contour integral from the MT and DOS contributions are proportional to finite-difference derivatives of the particle-hole bubble with respect to $p_0 = \iu \Omega_n$.
Since
\begin{equation}
    \frac{\abs{p_0}-\abs{p_0-q_0}}{q_0} = \frac{1}{\iu} \begin{cases}
        \displaystyle \frac{2 \Omega_n - \Omega_m}{\Omega_m} & 0 \leq \Omega_n < \Omega_m \\ 
        \sign{\Omega_n} &\text{otherwise},
    \end{cases}
\end{equation}
we see that the MT+DOS vertices contain a contribution similar to the AL diagram $\sim 1/(\iu p)$, but also another contribution $\sim \Omega_n/(\iu\Omega_m) \times 1/p$ for the range of Matsubara frequencies $0 < \Omega_n < \Omega_m $.
Together with the leading part of the AL vertices, the former represents the contribution to the fluctuation spin conductivity, denoted by $\sigma_{z}^{\mathrm{fluc}}(\omega)$.
The latter represents the contribution to the quasiparticle part of the spin conductivity, denoted by $\sigma_{z}^{\mathrm{qp}}(\omega)$.
In the following we compute each of these contributions separately.

\subsection{Fluctuation spin conductivity}
First we compute the contributions to spin transport from the spin fluctuations themselves. 
Here we use that $V_1^{R}(z,p) V_2^{R}(z,p) \sim - 1/p^2$ for the sum of all the diagrams, i.e., it is a constant function of $z$.
The regular part of the longitudinal spin conductivity is given by 
\begin{equation}
    \Re \sigma_{z}(\omega) \equiv  \Re \sigma_{z}^{\parallel}(\omega) = -\mathrm{P} \frac{1}{\omega} \Im[Q^{ii}_{\mathrm{fluc}}(\omega + \iu 0,\v{0})].
\end{equation}
By taking the imaginary part of Eq.~\eqref{eq:J_final} we find that
\begin{align}\label{eq:sigma_integral_supp}
    \Re \sigma_{z}^{\mathrm{fluc}}(\omega) \approx \frac{C e_{*}^2}{\omega} &\int \frac{\D^3 p}{(2\pi)^3} \int \frac{\D z}{2\pi}  V_1^{R}(z,p) V_2^{R}(z,p) \left[ \coth\left( \frac{z}{2 T}\right) - \coth\left( \frac{z-\omega}{2 T}\right)  \right] \Im D^{R}(z,\v{p}) \Im D^{R}(z-\omega,\v{p}),
\end{align}
where $C$ is a positive numerical prefactor.  
Using the general form of $D(\omega,\v{p})$ close to a $z=3$ QCP 
\begin{equation}
     D^{R}(z, \v{p}) = \frac{D_0}{\hat{\v{p}}^2 + \xi^{-2} - \iu \gamma z / (v_F \abs{\v{p}}) },
\end{equation}
and changing the integration variables to the dimensionless variables $\hat{p} = p/2k_F$, $\tilde{z} = \gamma z/4\epsilon_F$ and $\tilde{\omega} = \gamma \omega/4\epsilon_F$ we arrive at the following integral
\begin{equation}
    \Re \sigma_{z}^{\mathrm{fluc}}(\omega) \propto \int \D \hat{p} \hat{p}^2 \int \D \tilde{z} \frac{\coth(\alpha [\tilde{z} - \tilde{\omega}] ) - \coth(\alpha \tilde{z})}{\tilde{\omega}} \frac{\tilde{z} (\tilde{z}- \tilde{\omega})}{\left[  A^2(\hat{p}) + \tilde{z}^2 \right] \left[ A^2(\hat{p}) + (\tilde{z}-\tilde{\omega})^2 \right]},
\end{equation}
where $A(\hat{p}) \coloneqq \hat{p} (\hat{p}^2 + \xi^{-2})$ and $\alpha \coloneqq 2 \epsilon_F / (\gamma T)$.
After defining the new integration variables $\nu = \tilde{z} \xi^{3}$ and $\rho = \hat{p} \xi$ we can express the fluctuation spin conductivity in terms of a scaling function $\mathcal{F}(\varpi, \alpha\xi^{-3})$ as
\begin{equation}
    \Re \sigma_{z}^{\mathrm{fluc}}(\omega) \propto  T \xi^6 \mathcal{F}(\varpi, \alpha\xi^{-3}),
\end{equation}
where
\begin{equation}
    \mathcal{F}(\varpi, \alpha\xi^{-3}) = \int_{0}^{\infty} \D \rho \rho^2 \int_{-\infty}^{+\infty} \D \nu  \frac{\mathcal{A}(\nu,  \nu - \varpi; \alpha\xi^{-3}, \varpi)}{\left[ A^2(\rho) + \nu^2 \right] \left[ A^2(\rho) + (\nu - \varpi )^2 \right]},
\end{equation}
where $A(\rho) \coloneqq \rho (\rho^2 +1)$ and $\varpi \coloneqq \tilde{\omega} \xi^{3}$ and
\begin{equation}
    \mathcal{A}(\nu,  \nu - \varpi; \alpha\xi^{-3}, \varpi) = \alpha \xi^{-3} \nu (\nu - \varpi) \frac{\coth( \alpha\xi^{-3} \left[  \nu - \varpi \right] ) - \coth(\alpha \xi^{-3} \nu)}{\varpi}.
\end{equation}

\begin{figure}[t]
    \centering
    \includegraphics[width=0.5\linewidth]{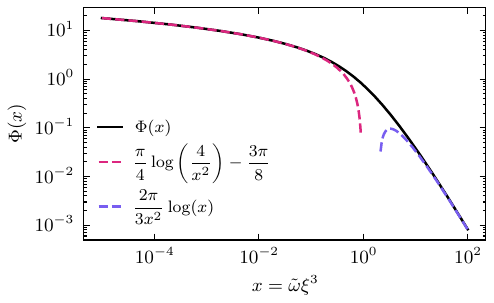}
    \caption{Scaling function $\Phi(x)$ that determines the frequency dependence of the conductivity in the limit $\alpha \xi^{-3}\ll 1$, computed from Eq.~\eqref{eq:Phi_def}.
    The figure also includes the asymptotic behaviors of $\Phi(x)$ as $x\ll 1$ and $x\gg1$.
    }
    \label{fig:scaling_function_supp}
\end{figure}

In the asymptotic limit $\alpha \xi^{-3} \ll 1$ we have
\begin{equation}
    \mathcal{A}(\nu,  \nu - \varpi; \alpha\xi^{-3}, \varpi) \to 1.
\end{equation}
This yields
\begin{align}
	\Re \sigma_{z}^{\mathrm{fluc}}(\omega) &\propto \alpha^{-1} \xi^6  \int_{0}^{\infty} \D \rho \rho^2 \int_{-\infty}^{+\infty} \D \nu  \frac{1}{\left[ A^2(\rho) + \nu^2 \right] \left[ A^2(\rho) + (\nu - \varpi )^2 \right]} \notag \\
	&= 2\pi \alpha^{-1} \xi^6  \int_{0}^{\infty} \D \rho \rho^2 \frac{1}{A(\rho) \left( 4 A^2(\rho) + \varpi^2 \right)}.
\end{align}
Hence, in this limit we find that the conductivity takes the form
\begin{equation}
	\Re \sigma_{z}^{\mathrm{fluc}}(\omega) \propto T \xi^6 \Phi( \tilde{\omega} \xi^3 ),
\end{equation}
where $\Phi(x)$ is the scaling function given by 
\begin{equation}\label{eq:Phi_def}
	\Phi(x) = 2\pi\int_{0}^{\infty} \D \rho \rho \frac{1}{\left( \rho^2 + 1 \right) \left( x^2 + 4 \rho^2 (\rho^2 + 1)^2 \right)} = \pi \int_{0}^{1} \D t \frac{t^2}{4 - 4t + x^2 t^3},
\end{equation}
where the second equality is obtained by introducing the change of variables $t = 1/(\rho^2+1)$.
In the limit where $x \ll 1$ we find that
\begin{equation}\label{eq:Phi_low_f}
	\Phi(x) \simeq \frac{\pi}{4} \log\left(\frac{4}{x^2}\right) - \frac{3\pi}{8},
\end{equation}
while for $x \gg 1$ we find that 
\begin{equation}
	\Phi(x) \simeq \frac{2\pi}{3x^2} \log(x).
\end{equation}
Plots of the scaling function and its asymptotic approximations are shown in Fig.~\ref{fig:scaling_function_supp}.
This implies that for $\alpha \xi^{-3} \ll 1$ the fluctuation conductivity takes the form
\begin{equation}
    \Re \sigma_{z}^{\mathrm{fluc}}(\omega) \sim T \xi^{6} \log\left( \frac{1}{\tilde{\omega} \xi^{3}} \right),
\end{equation}
in the low-frequency limit $\tilde{\omega}\xi^{3} \ll 1$.
Since this result holds for $\alpha \xi^{-3} \ll 1$ it applies to the quantum critical regime (QC in  Fig. 1 of the main text).

\begin{figure}[t]
	\centering
	\includegraphics[width=0.5\textwidth]{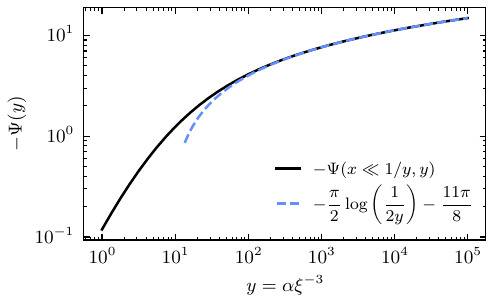}
	\caption{Scaling function $\Psi(y) \equiv \Psi(x\ll1/y,y) $ as a function of $y = \alpha \xi^{-3}$ together with its asymptotic behavior.}
	\label{fig:Psi_scaling}
\end{figure}

In  the opposite limit where $\alpha \xi^{-3} \gg 1$ we are in the Fermi liquid regime, which lies below the crossover line in Fig.~1 of the main text.
Here one cannot simplify the integral kernel $\mathcal{A}$ in the same way as before.
However, in the low-frequency limit we can write the conductivity as
\begin{equation}
    \Re \sigma_z^{\mathrm{fluc}}(\omega) \propto T \xi^6 \left[ \Phi(\tilde{\omega}\xi^3) + \Psi(\tilde{\omega}\xi^3, \alpha \xi^{-3}) \right],
\end{equation}
where 
\begin{equation}\label{eq:Psi_def}
    \Psi(x,y) = \int_{0}^{\infty} \D \rho \rho^2 \int_{-\infty}^{+\infty} \D \nu \left[ \frac{(\nu y)^2}{\sinh^2(\nu y)} - 1 \right] \frac{1}{\left[ A^2(\rho) + \nu^2 \right] \left[ A^2(\rho) + (\nu - x )^2 \right]}.
\end{equation}
When $x y \ll 1$ and $y \gg 1$ we find that
\begin{equation}
    \Psi(x \ll 1/y ,y) \simeq \frac{\pi}{2} \log\left(\frac{1}{2y}\right) + \frac{11 \pi}{8}.
\end{equation}
Hence, the logarithms of $\Phi$ and $\Psi$ combine and the leading logarithmic singularity in the Fermi liquid regime is instead
\begin{equation}
    \Re \sigma_{z}^{\mathrm{fluc}}(\omega) \sim T \xi^{6} \log\left( \frac{T}{\omega} \right),
\end{equation}
in the low-frequency limit $\tilde{\omega}\xi^{3} \ll 1$.
A plot of the function in Eq.~\eqref{eq:Psi_def} is shown in Fig.~\ref{fig:Psi_scaling}.

The log divergence of Eq.~\eqref{eq:Phi_low_f} as $\omega\to0$ derives from the infinite lifetime of the Landau-damped spin fluctuations in the IR limit.
Therefore, the strict $\omega\to0$ limit of the fluctuation part of the conductivity is inaccessible in this calculation due to effects not included in the model that would give these modes a finite lifetime.

\subsection{Quasiparticle contribution to the spin conductivity}
Let us now turn to the quasiparticle contribution to the spin conductivity, $\Re\sigma_{z}^{\mathrm{qp}}(\omega)$.
Since the additional factor of the vertex $\sim \Omega_n/(\iu\Omega_m)$ arises only in the regime $0 < \Omega_n < \Omega_m$ this affects only the terms involving $F^{RA}$ of Eq.~\eqref{eq:J_final}.
This implies that the conductivity now takes the form
\begin{equation}
    \Re \sigma_{z}^{\mathrm{qp}}(\omega) \approx \frac{C_2 e_{*}^2}{\omega^3} \int \frac{\D^3 p}{(2\pi)^3} \frac{1}{p^2}\int \frac{\D z}{2\pi} z^2 \left[ \coth\left( \frac{z-\omega}{2 T}\right) - \coth\left( \frac{z}{2 T}\right)  \right] \Re\left[ D^{R}(z,\v{p}) D^{A}(z-\omega,\v{p}) \right],
\end{equation}
where $C_2$ is a positive numerical prefactor. 
This contribution takes the form of the $1/\omega^2$ tail of the Drude contribution to the spin conductivity \cite{Kim-Lee-GaugeinvariantResponseFunctions-1994},
\begin{equation}
    \Re \sigma_{z}^{\mathrm{qp}}(\omega) \propto \frac{\Gamma}{\Gamma^2 + \omega^2} \approx \frac{\Gamma}{\omega^2},
\end{equation}
where the scattering rate $\Gamma = 1/\tau$ is computed by performing the $\omega \to 0$ limit of the integral that multiplies the $1/\omega^2$ tail.
This integral reads
\begin{equation}\label{eq:Gamma_expr}
    \Gamma \propto \alpha \int_{0}^{\infty} \D p p^2 \int_{-\infty}^{+\infty} \D z \frac{z^2}{\sinh^2(z/(2T))} \frac{1}{p^2} \abs{D^{R}(z,p)}^2 \propto \alpha^{-1} \int_{0}^{\infty} \D \rho \rho^2 \int_{-\infty}^{+\infty} \D \nu \frac{(\alpha\xi^{-3} \nu)^2}{\sinh^2(\alpha\xi^{-3} \nu)} \frac{1}{A^2(\rho) + \nu^2}.
\end{equation}

In the limit that $\alpha \xi^{-3} \ll 1$ we find that 
\begin{equation}
    \Gamma \propto \frac{\gamma T}{2 \epsilon_F} \log\left(\frac{\gamma T}{2\epsilon_F} \xi^{3}\right).
\end{equation}
Approaching the QCP from above, one has $\xi^{2} \sim \alpha^{2}$, which means that
\begin{equation}
    \Gamma \propto \frac{\gamma T}{2 \epsilon_F} \log\left(\frac{2\epsilon_F}{\gamma T}\right).
\end{equation}
This rate vanishes as one approaches the QCP.
The expression for $\Gamma$ in Eq.~\eqref{eq:Gamma_expr} is analogous to Eq.~(5) of Ref.~\cite{Duine-Vignale-SpinDragUltracold-2010}.
The spin-drag scattering rate computed by Ref.~\cite{Duine-Vignale-SpinDragUltracold-2010} treats the scattering processes as spin conserving, leading to an additional factor of $p^2$ in the scattering rate integral.
This implies that $\Gamma_{\mathrm{sd}} \propto T^{5/3}$ as the QCP is approached.
Since our system includes only transverse spin fluctuations this reduction of the relaxation rate does not take place.
This is manifested by the fact that the exact sum of the MT and DOS diagrams in our case produces exactly the vertex $\Upsilon^{ij}(p)$ which, after performing the angular average, is $\propto v_F^2$ and not $(p/m)^2$ (cf. Eq.~\eqref{eq:im_Upsilon_tilde}).
In the Fermi liquid regime $\alpha \xi^{-3} \gg 1$ we find $\Gamma \propto \xi^{-3}/(\alpha \xi^{-3})^2$.

\section{Fluctuation corrections to the electrical conductivity}\label{sec:electrical}

In contrast to the electrical conductivity, the spin conductivity is not protected by the conservation of momentum in the ultraclean single-band system.
This is consistent with the fact that the contributions to the spin conductivity from the Gaussian-fluctuation diagrams do not cancel.
To corroborate this result we now demonstrate that the fluctuation corrections to the electrical conductivity vanish within the same formalism.

The kernel $K^{\mu\nu}(q)$ defining the electrical current in linear response is given by
\begin{equation}
    J^{\mu}(q) = K^{\mu\nu}(q) A_{\nu}(q),
\end{equation}
where $A_{\mu}$ now denotes the real electromagnetic gauge field.
To compute this we follow the same approach as before, but replacing the matrix $\sigma^{z}/2$ in Eq.~\eqref{eq:current_vertex} by the identity matrix and the factor of $1/4$ in Eq.~\eqref{eq:dia_current_vertex} by $1$.
The charge now corresponds to the real electric charge. 
Notably, these replacements give rise to differences in the spin traces.
In particular, the two terms in Eq.~\eqref{eq:AL_vertex_realspace} now come with the same relative sign, and the sign of the MT diagram is changed.
Specifically, the AL vertex reads
\begin{align}\label{eq:AL_electrical}
		\Lambda^{\mu}(p-q,p) = \frac{e}{4}\frac{1}{\beta V} \sum_{k}&G_{0}(k) \gamma^{\mu}(k,k-q) G_{0}(k-q) \left[ G_{0}(p+k-q) + G_{0}(k-p)\right].
\end{align}
The MT and DOS diagrams read
\begin{align}\label{eq:MT_electrical}
	K_{\mathrm{MT}}^{\mu\nu}(q) = -\frac{e^2}{2} \frac{1}{(\beta V)^2} \sum_{kp} D(p)  G_{0}(k-q) \gamma^{\mu}(k-q,k) G_{0}(k) G_{0}(k+p) \gamma^{\nu}(k+p,k+p-q) G_{0}(k+p-q). 
\end{align}
and
\begin{align}\label{eq:DOS_electrical}
\begin{split}
    K_{\mathrm{DOS}}^{\mu\nu}(q) = -\frac{e^2}{2} \frac{1}{(\beta V)^2} \sum_{kp}   D(p) [G_{0}(k)]^2 G_{0}(k+p) \bigg[ &\gamma^{\mu}(k,k+q) G_{0}(k+q) \gamma^{\nu}(k+q,k)  \\
    + &\gamma^{\mu}(k,k-q) G_{0}(k-q) \gamma^{\nu}(k-q,k) \bigg]. 
\end{split} 
\end{align}

We now evaluate these diagrams at $\v{q} = 0$ to compute the fluctuation corrections to the electrical conductivity \cite{Lundemo-Sudbo-FluctuationConductivityUltraclean-2026b,Boyack-Boyack-RestoringGaugeInvariance-2018}.
We consider first the sum of the MT and DOS diagrams.
For the spin conductivity we found that these diagrams added up to yield Eq.~\eqref{eq:sum_MT_DOS}.
Since these diagrams now come with the same sign the result is instead
\begin{align}
	K^{ij}_{\mathrm{MT}+\mathrm{DOS}}(q_0,\v{0}) = -\frac{e^2}{2 q_0^2  } \frac{1}{(\beta V)^2}\sum_{k p} \gamma^{i}(k,k) \gamma^{j}(p,p) G_{0}(k) G_{0}(k+p)  \left[ 2D(p) - D(p-q_0) - D(p+q_0) \right].
\end{align}
Now, observe that the sum over $k$ corresponds to a particle-hole bubble with a single current vertex insertion
\begin{equation}
    \Pi^i(p) = \frac{1}{\beta V} \sum_{k} \gamma^i(k,k) G_0(k) G_0(k+p).
\end{equation}
However, since this is contracted with an odd function in $p$ only the odd part of this bubble enters.
Recall that for the spin conductivity, the even part of this bubble appeared in the AL vertex in Eq.~\eqref{eq:AL_upsilon_form}.
Some straightforward manipulations like those used in Ref.~\cite{Lundemo-Sudbo-FluctuationConductivityUltraclean-2026b} now yields
\begin{equation}\label{eq:odd_part_Upsilon}
    \frac{\Pi^i(p) - \Pi^{i}(-p)}{2} = - \frac{1}{2} \gamma^{i}(p,p) \Pi(p).
\end{equation}
Inserting this into the sum of the MT and DOS diagrams yields
\begin{align}\label{eq:MTDOS_electrical}
	K^{ij}_{\mathrm{MT}+\mathrm{DOS}}(q_0,\v{0}) &=\frac{e^2}{4 q_0^2 } \frac{1}{\beta V}\sum_{ p} \gamma^{i}(p,p) \gamma^{j}(p,p) \Pi(p) \left[ 2D(p) - D(p-q_0) - D(p+q_0) \right] \notag \\
	=& \frac{e^2}{4 q_0^2} \frac{1}{\beta V} \sum_{p} \gamma^{i}(p,p) \gamma^{j}(p,p) \left[ D(p) - D(p-q_0) \right] \left[ \Pi(p) - \Pi(p-q_0) \right] \notag \\
	=&\frac{e^2}{\beta V}  \sum_{p} \gamma^{i}(p,p) \gamma^{j}(p,p) D(p) \left[ \frac{D^{-1}(p-q_0) - D^{-1}(p)}{q_0} \right] D(p-q_{0}) \left[ \frac{D^{-1}(p-q_0) - D^{-1}(p)}{q_0} \right].
\end{align}

Let us now consider the AL vertex in the uniform limit.
Making use of Eq.~\eqref{eq:partial_fraction_identity} we find 
\begin{align}
	\Lambda^{i}(p-q_0, p) &= \frac{e}{4 \beta V} \frac{1}{q_0} \sum_{k} \gamma^{i}(k,k) \left[ G_0(k - q_0) - G_0(k) \right] \left[ G_0(k+p-q_0) + G_0(k-p)\right] \notag \\
	&= \frac{e}{4 q_0} \left[ \Pi^{i}(p) + \Pi^{i}(-p+q_0) - \Pi^{i}(p-q_0) - \Pi^{i}(-p) \right].
\end{align}
Hence, we see that the AL vertex is the finite-difference derivative of the odd part of the particle-hole bubble with a single current vertex insertion.
With the help of Eq.~\eqref{eq:odd_part_Upsilon} we find
\begin{align}
	\Lambda^{i}(p-q_0, p) = - \frac{e}{4 q_0} \gamma^{i}(p,p) \left[ \Pi(p) - \Pi(p-q_0) \right] = \frac{e}{q_0} \gamma^{i}(p,p) \left[ D^{-1}(p) - D^{-1}(p-q_0) \right].
\end{align}
The complete AL diagram now reads
\begin{align}\label{eq:AL_electrical_complete}
	K^{ij}_{\mathrm{AL}}(q_0,\v{0}) = - \frac{e^2}{\beta V} \sum_{p} D(p) \gamma^i(p,p) \left[ \frac{D^{-1}(p) - D^{-1}(p-q_0)}{q_0}\right]  D(p-q_0) \gamma^{j}(p,p)\left[ \frac{D^{-1}(p) - D^{-1}(p-q_0)}{q_0}\right].
\end{align}
The sum of the fluctuation diagrams that contribute to the regular part of the electrical conductivity in Eq.~\eqref{eq:MTDOS_electrical} and Eq.~\eqref{eq:AL_electrical_complete} vanishes.
This result reflects the conservation of momentum in the ultraclean system.
A similar cancellation in the theory of superconducting fluctuations was found in Refs.~\cite{Boyack-Boyack-RestoringGaugeInvariance-2018,Lundemo-Sudbo-FluctuationConductivityUltraclean-2026b}.

%\newpage

%\bibliography{spin_conductivity_references}
%\end{document}
    
\end{document}